\documentclass[a4paper,11pt]{article}
\pdfoutput=1 
\usepackage{graphicx}
\usepackage{float}
\usepackage{mciteplus}
\usepackage{dirtytalk}

\usepackage{jheppub} 

\usepackage[T1]{fontenc} 

\renewcommand{\[}{\begin{equation}}
\renewcommand{\]}{\end{equation}}
\def\beq{\begin{equation}}
\def\eeq{\end{equation}}
\newcommand{\be}{\begin{eqnarray}}
\newcommand{\ee}{\end{eqnarray}}

\renewcommand{\texttt}{{}}

\def\bs{\begin{subequations}}
\def\es{\end{subequations}}

\def\Cc{\mathcal{C}}
\def\Dc{\mathcal{D}}

\def\Fc{\mathcal{F}}

\def\Kc{\mathcal{K}}
\def\Lc{\mathcal{L}}
\def\Mc{\mathcal{M}}

\def\Oc{\mathcal{O}}
\def\Pc{\mathcal{P}}

\def\Rc{\mathcal{R}}

\def\Wc{\mathcal{W}}

\newcommand{\tia}[1]{}

\newcommand{\bea}{\begin{eqnarray}}
\newcommand{\eea}{\end{eqnarray}}
\newcommand{\beas}{\begin{eqnarray*}}
\newcommand{\eeas}{\end{eqnarray*}}
\newcommand{\bal}{\begin{aligned}}
\newcommand{\eal}{\end{aligned}}

\def\({\left(}
\def\){\right)}

\newcommand{\LF}{\left(}
\newcommand{\RF}{\right)}
\newcommand{\LT}{\left[}
\newcommand{\RT}{\right]}

\newcommand{\D}{\mathcal{D}}

\newcommand{\pd}{\partial}

\newcommand{\const}{\mathrm{const}}

\title{Generalized non-local $R^2$-like inflation}

\author[a,b,f]{Alexey S. Koshelev,}

\author[c,d]{K. Sravan Kumar,}

\author[e,f]{Alexei A. Starobinsky}

\affiliation[a~]{
School of Physical Science and Technology, ShanghaiTech University, 201210 Shanghai, China
}
\affiliation[b~]{
	Departamento de F\'isica, Centro de Matem\'atica e Aplica\c{c}oes (CMA-UBI),
	Universidade da Beira Interior, 6200 Covilh\~a, Portugal }
\affiliation[c~]{Department of Physics, Tokyo Institute of Technology, 2-12-1 Ookayama, Meguro-ku, Tokyo 152-8551, Japan}
\affiliation[d~]{ Institute of Cosmology \& Gravitation,
	University of Portsmouth,
	Dennis Sciama Building, Burnaby Road,
	Portsmouth, PO1 3FX, United Kingdom}
\affiliation[e~]{
	L. D. Landau Institute for Theoretical Physics RAS, Chernogolovka, Moscow region 142432,
	Russian Federation}
\affiliation[f~]{Kazan Federal University, Kazan 420008, Republic of Tatarstan, Russian
	Federation}

\emailAdd{askoshelev@shanghaitech.edu.cn}
\emailAdd{sravan.kumar@port.ac.uk}
\emailAdd{alstar@landau.ac.ru}

\abstract{The $R^2$ inflation which is an extension of general relativity (GR) by quadratic scalar curvature introduces a quasi-de Sitter expansion of the early Universe governed by Ricci scalar being an eigenmode of d'Alembertian operator. 
	In this paper, we derive a most general theory of gravity admitting $R^2$ inflationary solution which turned out to be higher curvature non-local extension of GR.  We study in detail inflationary perturbations in this theory and analyse the structure of form-factors that leads to a massive scalar (scalaron) and massless tensor degrees of freedom. We argue that the theory contains only finite number of free parameters which can be fixed by cosmological observations. 
	We derive predictions of our generalized non-local $R^2$-like inflation and obtain the scalar spectral index $n_s\approx 1-\frac{2}{N}$ and any value of the tensor-to-scalar ratio $r<0.036$. In this theory, tensor spectral index can be either positive or negative $n_t\lessgtr 0$  and the well-known consistency relation $r = -8n_t$ is violated in a non-trivial way.  We also compute running of the tensor spectral index and discuss observational implications to distinguish this model from several classes of scalar field models of inflation. 
	These predictions allow us to probe the nature of quantum gravity in the scope of future CMB and gravitational wave observations. Finally we comment on how the features of generalized non-local $R^2$-like inflation cannot be captured by established notions of the so-called effective field theory of single field inflation and how we must redefine the way we pursue  inflationary cosmology. }

\arxivnumber{arXiv:2209.02515}

\makeatother

\begin{document}
	
\maketitle

\section{Introduction}

Ever since the first release of the Planck CMB data $R^2$ inflation, or Starobinsky inflation, based on modification of general relativity (GR) by quadratic scalar curvature (which we call shortly as $R^2$ gravity)  has became a cue for understanding the physics of early Universe \cite{Starobinsky:1980te,Starobinsky:1981vz,Starobinsky:1983zz,Vilenkin:1985md,Mijic:1986iv,Maeda:1987xf,Akrami:2018odb} because of its success with the predictions of a quasi-flat (scale invariant) scalar power spectrum of primordial scalar perturbations $\Pc_{\Rc}\sim 2.1\times 10^{-9}$ with the small tilt $n_s-1=-\frac{2}{N}$ and primordial gravitational waves given by the tensor-scalar ratio $r=\frac{12}{N^2}=3(1-n_s)^2$ (where $N=50-60$ e-folds before the end of inflation, depending on the duration of a post-inflationary stage of creation and thermalization of known matter), obtained by unambiguously fixing the mass of the scalaron as $M\approx 1.3\times 10^{-5}\,M_p$ in the units of reduced Planck mass ($M_p$) from the measured amplitude of $\Pc_{\Rc}$. The emergence of $R^2$ inflation from the semi-classical modification of gravity due to matter quantum fields clearly points out the importance of fundamental physics and the first principles in the construction of successful  cosmological models. Along with the observational success, $R^2$ gravity is also an important step beyond general relativity (GR) towards quantum gravity \cite{Stelle:1976gc}. In this spirit, $R^2$-inflation has been successfully embedded in a geometric (\'a la Jordan frame) construction of supergravity \cite{Ketov:2012jt} and also recently in an ultraviolet (UV) complete framework of analytic non-local gravity \cite{Koshelev:2016xqb,Koshelev:2017tvv,Koshelev:2020foq,Koshelev:2020xby}. Despite these developments especially the later one that we explore further in this paper, several realizations of Starobinsky-like inflatonary scenarios have been emerged in recent years in the context of UV-complete frameworks such as in string theory and SUGRA \cite{Ellis:2013nxa,Kehagias:2013mya,Linde:2014nna}. The important aspect of all these several generalizations is that they modify the scalar power spectrum which results in modification of the key inflationary observable - the tensor-to-scalar ratio which is bounded above from the latest BICEP/Keck Array CMB data $r<0.036$ \cite{BICEP:2021xfz}, while the prediction of scalar tilt $n_s\approx 1-\frac{2}{N}$ being almost unchanged in agreement with the Planck result \cite{Akrami:2018odb}.

Apart from the scalar field models of inflation which predominantly attempts to modify the scalar power spectrum,  there is another possibility of changing the key parameter tensor-to-scalar ratio $r$ by modifying the gravitational power spectrum. Such a possibility has been recently found in geometrical modifications of GR, in particular in an ultraviolet (UV) non-local generalization of $R^2$-inflation.
Such a non-local gravity theory has emerged with several motivations of UV completing $R^2$ gravity  (see \cite{Briscese:2013lna,Craps:2014wga,Koshelev:2016xqb,Koshelev:2017tvv,Koshelev:2020xby} and the references therein\footnote{See also \cite{Modesto:2021ief,Modesto:2021soh,Modesto:2022asj}
for other non-local gravity models not containing an inflationary stage.}). Recently found non-local extension of $R^2$ gravity leads to testable predictions, especially with respect to the single field consistency relation  $r=-8n_t$ which gets violated solely due to the presence of non-locality. Here in this paper, we establish $R^2$-like inflation in an analytic non-local gravity construction including all possible higher curvatures {(that are relevant up to the inflationary energy scales)} which go beyond the previously studied model \cite{Koshelev:2017tvv,SravanKumar:2018dlo,Koshelev:2020foq}. Non-local higher curvature terms are motivated from quantum gravity point of view which can be witnessed from the studies of trace-anomaly driven effective theory of quantum gravity  \cite{Barvinsky:1990up,Barvinsky:1994cg,Barvinsky:1994hw}  and from asymptotic safety (AS) approach to quantum gravity  \cite{Bonanno:2017pkg,Eichhorn:2018yfc,Bosma:2019aiu}. According to the AS approach, a generic (effective) quantum gravity action must contain all the higher curvatures and infinite derivatives. In this paper, we study  cosmological implications of non-local higher curvature terms in the view of $R^2$-like inflationary paradigm. In a nutshell, our primary goal is to expose all aspects of higher curvature gravity embodied with non-locality in the form of analytic infinite derivatives and higher curvature terms. We present features of this theory with respect to the almost scale invariant scalar power spectrum and the tensor power spectrum that can have non-trivial scale dependence.  

The paper is organized as follows. 
In \textbf{Sec.~\ref{sec:results}} we provide a concise summary of results obtained in this paper which will be discussed  in detail in the later sections of the paper.  We highlight all the new features of our geometric framework of non-local $R^2$-like inflation in the scope of upcoming ground based and space based cosmological and gravitational wave observations. 
In \textbf{Sec.~\ref{sec:hcmodel}} we present a most general action for $R^2$-like inflation and explain how the action turns out to be non-local in nature.  We demonstrate how the theory, despite containing infinite derivative and infinite curvature terms (written as the so-called form-factors), can be formulated with finite parameter space using physically motivated conditions for the theory to be ghost-free in the Minkowski and quasi-de Sitter (dS) limits.  
In \textbf{Sec.~\ref{sec:pwnaid}} we derive the scalar and tensor power spectrum predictions of our general higher curvature non-local gravity exploiting the results derived in \cite{Koshelev:2016xqb,Koshelev:2017tvv,Koshelev:2020foq}. We discuss the structure of form-factors and their relation to the degrees of freedom in the theory. We derive predictions for  $\LF n_s,\,r,\,\frac{dn_t}{d\ln k},\,\frac{d^2n_t}{d\ln k^2} \RF$ in this model and compare them to the latest observational bounds. 
In \textbf{Sec.~\ref{sec:eftcom}} we provide an extensive discussion of distinguishing the features of generalized non-local $R^2$-like inflation against the effective field theory (EFT) of single field inflation. We present how generalized non-local $R^2$-like inflation reshapes our understanding of early Universe cosmology. 
In \textbf{Sec.~\ref{sec:Conc}} we provide a brief outlook to go beyond what has been studied in this paper. Also, we will comment on implications of our study in the context of quantum gravity. 
In \textbf{Appendix~\ref{App:LR2in}} we discuss several models of $f(R)$ gravity and the prominence of $R^2$ inflation in explaining the inflationary observables. 
In \textbf{Appendix~\ref{App:EoM}} we derive the equations of motion of the most general gravity action for $R^2$-like inflation we constructed in Sec.~\ref{sec:hcmodel}. In \textbf{Appendix~\ref{App:pEoM}} we study cosmological perturbations of theory obtained in Sec.~\ref{sec:hcmodel} and analyze the degrees of freedom and the ghost-free conditions for different form-factors. 

Notations: we use the metric signature $\LF -,\,+,\,+,\,+ \RF$, overdot and $^\prime$ denotes derivative with respect to cosmic time (t) and conformal time ($ \tau $) respectively, $^\dagger$,\,$^{\dagger\dagger}$,\,$^{\dagger\dagger\dagger}$ to denote first, second and third derivative with respect to the argument. $^{(n)}$ denotes n-th order perturbation, overbar denotes background values for flat Friedmann-Lema\^itre-Robertson-Walker (FLRW) space-time, 4-dimensional indices are labeled by small Greek letters and three dimensional quantities are denoted by $i,j =1,2,3$. We also set $\hbar=c=1$ and $M_p=1/\sqrt{8\pi G}$ is the reduced Planck mass. Throughout this paper we use \say{$\approx$} to denote dS approximation, subscripts \say{$_{(s)},\,_{(h)}$} to indicate scalar and tensor parts of the perturbations and subscript \say{$_{\rm dS}$} refers to the quantities evaluated in the quasi-dS (slow-roll) approximation. Everywhere we perform computations in the leading order of the slow-roll approximation. Whenever we discuss scalar field EFT models of inflation, we use the notation for slow-roll parameters, Hubble parameter with subscript \say{E} to distinguish them from the quantities defined in the Jordan frame. We use the letter \say{O} to denote the order of magnitude and \say{$\Oc$} to denote an arbitrary operator. 

\section{Summary of our results}
\label{sec:results}
Apart from discovering a new era in the remote past of our Universe, one of the main goals of inflationary cosmology is to look for signatures of high energy physics including gravity through CMB, astrophysical, and gravitational wave observations. A Plethora of inflationary mechanisms have been developed in the last decades in order to derive predictions of physics at UV energy and space-time curvature in terms of the following set of observables related to two-point correlations \cite{Martin:2013tda}
\begin{equation}
	\Bigg\{n_s,\, r,\, n_t \Bigg\} 
	\label{predictions}
\end{equation}
The latest Planck data with their result on the scalar spectral index $n_s\approx 1-\frac{2}{N}$ and constraint on its running $\frac{dn_s}{d\ln k }$ strongly favour $R^2$ inflation where the inflationary stage is a phase of quasi-dS evolution with the Ricci scalar R being the eigenmode of the d'Alembertian operator \eqref{mR2in} which in the Einstein frame is understood as inflaton rolling down the plateau-like potential (or the Starobinsky potential) \cite{Starobinsky:1980te,Mukhanov:1990me}. $R^2$ inflation interpolates the regime of higher curvature dominance when $M^2\ll R\ll M_p^2$ (where $M \approx 1.3\times 10^{-5} M_p$ is the restmass of the propagating scalar in $f(R)$ gravity dubbed scalaron) to the matter dominated phase $|R|\ll M^2$ producing graceful exit from inflation, and when to the radiation dominated stage after the  decay of scalaron into other elementary particles and antiparticles and their final thermalization. The success of $R^2$ inflation puts stringent constraints on coefficients of additional higher order terms in the action like $R^n$. 

Before we attempt to understand what quantum gravity is, it is vital to understand how one can constructively write an action beyond GR. It is trivial to think that a generic action for quantum gravity (that is metric compatible) contains all possible curvature invariants that one can write using the Ricci scalar, Ricci tensor, Weyl tensor or Riemann tensor and an infinite number of their covariant derivatives. But it is non-trivial to frame out the rules to obtain an action that is apt for quantum gravity and cosmological application. Inspired from the attempts of formulating a generic action of quantum gravity with all possible curvature invariant terms \cite{Barvinsky:1990up,Barvinsky:1994cg,Barvinsky:1994hw} and several other approaches of quantum gravity \cite{Koshelev:2020xby}, here in this paper we took a phenomenological approach and formulated a most general action \eqref{NAID} that one can write to have $R^2$-like inflation dictated by the eigenvalue equation  \eqref{mR2in} as its particular solution. We found that our generalized action \eqref{NAID} contains infinite order curvature terms and infinite order of their derivatives, thus becoming a non-local theory of gravity. Furthermore, we realized that the earlier studied $R^2$-like inflation in quadratic curvature non-local theory \cite{Koshelev:2016xqb,Koshelev:2017tvv,Koshelev:2020foq,Koshelev:2020xby} is an effective description of our more fundamental action \eqref{NAID}. The crux of the paper is about inflationary predictions of our gravity action \eqref{NAID} for the following inflationary observables related to two-point scalar and tensor correlations, scalar and tensor spectral tilts and their running
\begin{equation}
	\Bigg\{n_s,\, \frac{dn_s}{d\ln k},\,r,\, n_t,\, \frac{dn_t}{d\ln k},\,  \frac{d^2n_t}{d\ln k^2}\Bigg\}
	\label{inobservables}
\end{equation}
\begin{figure}[h]
	\centering
	\includegraphics[width=0.7\linewidth]{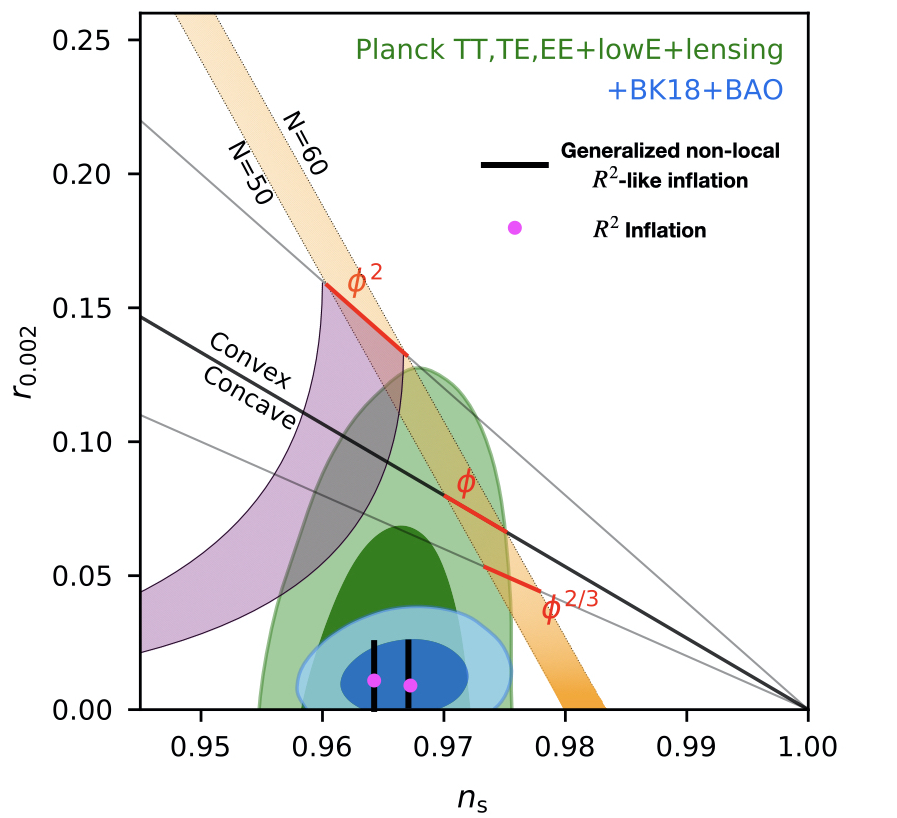}
	\caption{In this plot we confront the predictions of generalized non-local $R^2$-like inflation  against the latest BICEP/Keck array analysis (depicting the ruled out monomial models of inflation and the natural inflation which is in purple colour)  \cite{BICEP:2021xfz} which has constrained the tensor-to-scalar ratio as $r<0.036$ and the spectral index as $n_s=0.9649\pm 0.0042$. The black colour vertical line from the left represents $\LF n_s,\,r\RF = \LF 0.964,\, < 0.036 \RF$ for $N=55$ whereas the one on the right represents $\LF n_s,\,r\RF = \LF 0.967,\, < 0.036 \RF$ for $N=60$. }
	\label{nsr}
\end{figure}
We have shown that the power spectrum predictions of our generalized action remain the same as in the earlier studied model of $R^2$-like inflation in non-local gravity \cite{Koshelev:2020foq} with the theoretically consistent choice of form-factors which do not explicitly depend on the background curvature.  Especially, we have established in detail the structure of form-factors and degrees of freedom of the theory (See Sec.~\ref{sec:pwnaid} along with Appendix~\ref{App:pEoM} for details). The predictions of $\LF n_s,\,r \RF$ are depicted in Fig.~\ref{nsr} against the latest bounds on the tensor-to-scalar ratio from BICEP2/Keck Array \cite{BICEP:2021xfz}. In this theory we modify the tensor power spectrum as in \eqref{pt}, whereas the scalar power spectrum remains nearly scale invariant, and also the scalar spectral index $n_s$ and its running $\frac{dn_s}{d\ln k}$ remain the same as in the local $R^2$ theory. 

Here are the key consequences of the modification of the tensor power spectrum which is due to terms of infinite order in powers of Ricci scalar, Weyl tensor squared and their covariant derivatives in \eqref{NAID}. 
\begin{itemize}
	\item Sound speed of tensor fluctuations remains Unity that is compatible with the latest bounds on the sound speed of gravitational waves from GW170817 \cite{baker2022measuring}. Also, the sound speed of scalar fluctuations in our model remains Unity. 
	\item Single field tensor consistency relation is modified, $r\neq -8n_t$, due to modification of the tensor tilt by non-local higher curvature effects.  
	\item Scale invariant nature of tensor power spectrum\footnote{In GR, scale invariance of the power spectrum of tensor perturbations generated during a metastable de Sitter (inflationary) stage in the early Universe was first derived in~\cite{Starobinsky:1979ty} without introducing a concrete model of inflation.} can be broken and the predictions of $\LF n_t,\, \frac{dn_t}{d\ln k} \RF$ can be an order of magnitude larger than in conventional EFT models of inflation (See Sec.~\ref{sec:pwnaid} and the discussion in Sec.~\ref{sec:eftcom}). 
	\item The tensor tilt can be positive as well as negative. Especially, it is possible to achieve blue-tilt\footnote{Alternative theories of inflation such as string gas cosmology can have blue tilt, too \cite{Wang:2014kqa}, but such frameworks predict undetectably small non-Gaussianities. In our case, we can have blue tilt, but there can be relatively large non-Gaussianities \cite{Koshelev:2022bvg,KKS2} that is a distinguishable feature from string gas cosmology}  $n_t>0$ for any value of $r<0.036$. 
	\item In our framework, the Hubble parameter during inflation remains the same for any value of tensor-to-scalar ratio $r<0.036$. Therefore, the energy scale of inflation remains unchanged with changing the value of $r$. 
	\item We have shown in detail that our non-local gravity inflation cannot fall into any category of EFT of single field inflation \cite{Cheung:2007st,Cabass:2022avo} which is established as the unified framework of all single scalar field models of inflation.  
\end{itemize} 

All together we successfully establish in this paper that geometric modifications of $R^2$ inflation can lead to definite predictions for inflation which are in-line but still distinguishable from several classes of scalar field inflationary models. Thus, we open a new door for understanding and development of cosmic inflation.  Our study will be continued in the companion papers where we study primordial scalar and tensor non-Gaussianities \cite{Koshelev:2022bvg,KKS2}. All of the above predictions make the generalized non-local $R^2$-like inflation a viable target for the future experiments such as CMBS4 \cite{CMB-S4:2020lpa}, LiteBIRD \cite{LiteBIRD:2022cnt}, e-LISA \cite{Ricciardone:2016ddg} and CORE \cite{CORE:2016ymi} which are aimed for detection of primordial gravitational waves.

\section{Generalized $R^2$-like inflation in analytic non-local gravity}
\label{sec:hcmodel}

\subsection{Action of the model}

Given the success of $R^2$ inflation, any extension of it must preserve the values of the scalar power spectrum and its tilt which are accurately measured by the Planck data \cite{Akrami:2018odb}. Several finite derivative higher curvature  extensions of $R^2$ gravity have been studied in the literature (See Appendix~\ref{App:LR2in} for a detailed discussion) which all conclude that $R^2$-inflation is the consistent setup to describe current bounds on spectral index and its running for $N\sim 50-60$ e-foldings. In this Section, we present the most general gravity action mostly up to cubic in curvature terms which can be written such that the original inflationary background solution remains the exact one. The latter is known to form the eigenvalue problem\footnote{This is the trace equation of $R^2$ theory with the action
	$
	S^{\rm local}_{R+R^2} =\frac{1}{2} \int d^4x\sqrt{-g}\,( M_p^2R+f_0R^2)\,
	$ with $f_0=\frac{M_p^2}{6M^2}$.}
\begin{equation}
	\square R= M^2R\, 
	\label{mR2in}
\end{equation}
A tedious but straightforward calculation leads to the following gravity action possessing this property that contains infinite order of derivatives and in some places infinite order curvature terms 
\begin{equation}
	\begin{aligned}
		S_H^{\rm Non-local} 
 = 		 &\,\frac{1}{2}\int d^4x\sqrt{-g}\, \Bigg(M_{p}^2R +
		\Bigg[ R\Fc_R\LF \square_s \RF R   +  \LF \frac{M_p^2}{2\Mc_s^2}+ f_0 R_s \RF W_{\mu\nu\rho\sigma}{\mathcal{F}}_{W}\left(\square_{s},\, R_s \right)W^{\mu\nu\rho\sigma}  \\
		&+ \frac{f_0\lambda_c}{\Mc_s^2}\Lc_1\LF \square_{s} \RF R\, \Lc_2\LF \square_{s} \RF R\, \Lc_3\LF  \square_{s}\RF R \\
		& +\frac{f_0\lambda_R}{\Mc_s^2}\Dc_1\LF \square_s \RF R\Dc_2\LF \square_s\RF W^{\mu\nu\gamma\lambda}\Dc_3\LF \square_s \RF W_{\mu\nu\gamma\lambda}
		\\
		& +\frac{f_0\lambda_W}{\Mc_s^2}\Cc_1\LF \square_s \RF W_{\mu\nu\rho\sigma}\Cc_2\LF \square_s\RF W^{\mu\nu\gamma\lambda}\Cc_3\LF \square_s \RF W_{\gamma\lambda}^{\quad\rho\sigma}\Bigg]+\cdots\Bigg)\,,
		\label{NAID}
	\end{aligned}
\end{equation}
where $\cdots$ represent higher order curvature terms which do not contribute neither to two-point nor to 3-point correlations of inflationary fluctuations. In the companion papers \cite{Koshelev:2022bvg,KKS2} we further disclose details about these terms. Here $R,\,W_{\mu\nu\rho\sigma}$ are Ricci scalar and Weyl tensor respectively and $\square_s = \frac{\square}{\Mc_s^2}$ with $\square$ being the d'Alembertian operator. $\Mc_s$ is the scale of higher derivative corrections, $R_s=\frac{R}{\Mc_s^2}$,  and 
\begin{equation}
	\Bigg\{ \Fc_R\LF \square_s \RF,\,\Fc_W\LF \square_s,\, R_s\RF,\,\Lc_i\LF \square_s \RF,\,\Cc_i\LF \square_s \RF,\,\Dc_i\LF \square_s \RF\Bigg\}
	\label{formfactl}
\end{equation}
are analytic infinite derivative (AID) operators often called form-factors. We assume that $M^2\ll\Mc_s^2\ll M_p^2$ and {also, since the scope of \eqref{NAID} is only limited to the inflationary context,} we assume that $R\ll M_p^2$ throughout our analysis. Analyticity at zero curvature is essential to restore Einstein's gravity in IR. {We stress that \eqref{NAID} is structurally very different from the earlier studied versions of non-local gravity \cite{Koshelev:2017tvv,Koshelev:2016xqb}. Compared to the non-local gravity model studied in \cite{Koshelev:2017tvv,Koshelev:2016xqb}, the action \eqref{NAID} obviously contains higher curvature terms but the term $\LF \frac{M_p^2}{2\Mc_s^2} + f_0R_s\RF W_{\mu\nu\rho\sigma}\Fc_W\LF \square_s,\, R \RF W^{\mu\nu\rho\sigma}$, in particular, contains a combination of infinite curvature terms involving Ricci scalar which are constrained by the ghost-free structure of form-factor \eqref{formfinal} around the inflationary background that shall be explained in detail in the later part of the section. We should observe here that in the previous frameworks \cite{Koshelev:2017tvv,Koshelev:2016xqb} this $R$ dependence in the non-local Weyl square term was not present. However, the form-factor considered there does contain the particular dependence on the value of quasi-dS curvature during inflation (see Eq. (5.14) of \cite{Koshelev:2016xqb}). Therefore, our formulation of \eqref{NAID} gives a consistent extension of the non-local gravity theory studied in \cite{Koshelev:2017tvv,Koshelev:2016xqb}.  }


Let us elaborate on how action (\ref{NAID}) emerges.
In general, any higher curvature extension of GR is constructed through adding all possible curvature invariants involving the following three curvature tensors
\begin{equation}
	\mathbb{R} =\Bigg\{ R\quad R_{\mu\nu}\quad W_{\mu\nu\rho\sigma}\Bigg\}\,,
	\label{basis}
\end{equation}
where $\mathbb{R}$ denotes either Ricci scalar, or Ricci tensor, or the Weyl tensor. Since Riemann tensor can be expressed in terms of the basics \eqref{basis}, we can drop it from our construction of the action. Weyl tensor is a convenient choice for FLRW backgrounds compared to Riemann tensor.  
In the context of inflationary backgrounds, since $\bar{R}_{\mu\nu}\approx \frac{\bar{R}}{4}\,g_{\mu\nu}$, it is wise to eliminate terms in the action containing Ricci tensor to simplify our analysis. Moreover, our task here is to find a pure gravitational action compatible with \eqref{mR2in}. Therefore, we can resort for the moment to just terms containing the scalar curvature $R$ leaving the detailed analysis of the impact of the terms containing Ricci tensor for future studies.

Extending Einstein's gravity to a 4-derivative action, we end up with the Stelle gravity
\begin{equation}
	S_{4} = \frac {1}{2} \int d^4x \sqrt{-g}\,(M_p^2R+f_0R^2+f_{W0}W_{\mu\nu\rho\sigma}W^{\mu\nu\rho\sigma})\,,
	\label{Stelleg}
\end{equation}
and the trace equation here is identical to \eqref{mR2in}. Stelle gravity was found to be renormalizable \cite{Stelle:1977ry} but non-unitary because of the presence of the Weyl ghost\footnote{However at quantum level, the unitarity problem resolutions for Stelle gravity were elaborated by changing the Feynman prescription via the so-called Fakeon prescription or PT-symmetric quantization or by Renormalization Group approach by including matter couplings \cite{Anselmi:2017ygm,Mannheim:2021oat,Bender:2007wu,Salvio:2014soa,Salvio:2017qkx}. Motivated by these approaches, there have been studies of $R^2$-inflation and other local higher-derivative models and their respective UV-completions \cite{Anselmi:2020lpp,Salvio:2019agg,Salvio:2022mld,Liu:2022gun}.}. Even though non-unitary, Stelle gravity tells us that the higher curvature modification is the right approach towards quantum gravity due to manifest renormalizability of the model. Assuming the ghost mass to be heavy, i.e. $f_0\gg f_{W0}$,  Stelle gravity successfully leads to Starobinsky inflation \cite{Starobinsky:1980te}. Let us remind here that $f_0$ is very large indeed in the local $R^2$ inflation,
$f_0=\frac{N^2}{144\pi^2\Pc_{\Rc}}\approx 10^9$, that follows from the smallness of the primordial scalar perturbation spectrum $\Pc_{\Rc}$ only
($\Pc_{\Rc}(N)\propto N^2, N\gg 1$ for Starobinsky inflation, so $f_0$ is $N$-independent).

Later on it has become clear that by adding higher (infinite) derivative terms, an infinite derivative extension of Stelle gravity  \cite{Krasnikov:1987yj,Koshelev:2020xby} can lead to a ghost-free theory around Minkowski. {The action \eqref{AIDG} is found to have good UV complete properties such as super-renormalizability  \cite{Tomboulis:1997gg,Tomboulis:2015esa,Tomboulis:2015gfa,Modesto:2011kw,Biswas:2011ar,Koshelev:2016xqb,Biswas:2016egy,Koshelev:2017ebj,Calcagni:2014vxa}, and in some cases with more local higher curvature terms finite quantum gravity has also been known to be achieved by removing the finite loop divergences using the methods of power counting and $\beta$-function computations \cite{Modesto:2014lga,Koshelev:2017ebj,Koshelev:2016xqb,Briscese:2013lna,Modesto:2015foa,Modesto:2015lna,Calcagni:2014vxa}. However, still several open questions exist about unitarity of the theory at the non-perturbative level, behavior of scattering amplitudes and the uniqueness of higher derivative operators. To be more precise, the scattering amplitudes and the non-perturbative unitarity have been addressed in only a sub-class of non-local theories without the Weyl tensor terms  \cite{Dona:2015tra,Modesto:2021soh,Calcagni:2018gke,Calcagni:2018lyd}.  Therefore, we stress that the study of super-renormalizability or finiteness and also the non-perturbative unitarity of our generalized action \eqref{NAID} is a subject of future investigation. In our generalized non-local gravity \eqref{NAID} we have higher curvature non-local terms which are added for necessity to achieve a ghost-free behaviour in the quasi-dS limit. This is in the view of cosmological applicability through $R^2$-like inflation as a guiding principle for achieving a consistent non-local (quantum) gravity.}     
\begin{equation}
	S^{\rm Non-local}_q = \frac{1}{2}\int d^4x\sqrt{-g}\,(M_p^2R+R\Fc_R\LF \square_s \RF R+
	W_{\mu\nu\rho\sigma}\Fc_C\LF \square_s \RF W^{\mu\nu\rho\sigma})
	\label{AIDG}
\end{equation}
with the form-factors (see \cite{Biswas:2016egy,Koshelev:2016xqb} for more details)
\begin{equation}
	\begin{aligned}
		\Fc_R\LF \square_s \RF & = \sum_{n\geq0} f_n\square_s^n = f_0 M^2 \,\frac{1-\LF 1-\frac{\square}{M^2}  \RF e^{\gamma_S\LF \square_s \RF} }{\square}\\
		\Fc_C\LF  \square_s \RF &  = \sum_{n\geq0} f_{Wn} \square_s^n = \frac{M_p^2}{2}\, \frac{e^{\gamma_T\LF \square_s \RF}-1}{\square}\,. 
	\end{aligned}
	\label{formMin}
\end{equation}
where $\gamma_S\LF \square_s \RF,\,\gamma_T\LF \square_s\RF$ are entire functions, quite arbitrary in principle. The fact that these functions are entire and come in the exponent guarantees that extra degrees of freedom do not emerge. Note that finite number of derivatives will not kill ghosts in the spectrum, we need infinite number of derivatives! From the point of view of convergence of loop integrals, one want that form-factors $\Fc$ at least do not decay for large momenta (meaning that the propagator behaves at least like in Stelle gravity).

Though looking promising, this construction suffers from the inability to be healthy around any background but Minkowski. Indeed, it was shown explicitly in \cite{Biswas:2016etb} how the form-factors would look like around dS background (assuming explicit cosmological constant term in the action), and the crucial point is that Taylor coefficients of form-factors should implicitly depend on the background curvature. Even though one can do adjust the coefficient in the right way, it becomes possible for one given background (one given dS curvature) only. On the other hand, the very structure of the above action (\ref{AIDG}) does not assume an inclusion of extra curvature factors anywhere meaning that this model is yet incomplete.

In the present paper we want to break free out of the above problem by considering extra dependence on curvature. To start with and give a feeling of what we are looking for, we notice that since Weyl tensor is zero for any FLRW background, in principle we can include any curvature dependence inside the form-factor $\Fc_C\LF \square_s \RF$ without changing the propagator around Minkowski space-time (or more generically, as long as we are interested in FLRW background). In this regard, we can generalize the non-local Weyl square term in \eqref{AIDG} as
\begin{equation}
	S_{W^2}^{\rm Non-local}=  \frac{1}{2} \int d^4x\sqrt{-g}\Bigg[ \LF \frac{M_p^2}{2\Mc_s^2}+f_0\frac{R}{\Mc_s^2} \RF W_{\mu\nu\rho\sigma}\Fc_W\LF \square_s,\,R_s \RF W^{\mu\nu\rho\sigma}  \Bigg]
	\label{Genweyl}
\end{equation}
where $\Fc_W\LF \square_s,\,R_s \RF$ is a general function of d'Alembertian and Ricci scalar being analytic at zero values of its arguments. We can also include any arbitrary function of a generic Riemann tensor analytic at zero argument, but for the moment we restrict ourselves to the scalar curvature $R$ only. {We note that \eqref{Genweyl} is different from what was considered in \cite{SravanKumar:2018dlo}. First of all, the non-local Weyl tensor square term in \cite{SravanKumar:2018dlo} does not contain Ricci scalar-dependent factors but rather it contains terms involving a particular constant value of Ricci scalar (which is $\bar{R}_{\rm dS}$ in the notation of this paper). To be precise, the non-local Weyl tensor square term in action considered in \cite{SravanKumar:2018dlo} is $\LF \frac{M_p^2}{2\Mc_s^2}+f_0\frac{\bar{R}_{\rm dS}}{\Mc_s^2} \RF W_{\mu\nu\rho\sigma}\Fc_W\LF \square_s,\,\frac{\bar{R}_{\rm dS}}{\Mc_s^2} \RF W^{\mu\nu\rho\sigma}$ which matches with the one in \cite{Koshelev:2016xqb} in the approximation $M_p^2\ll f_0\bar{R}_{\rm dS}$. All these studies clearly signal that the earlier studied non-local gravity models \cite{Koshelev:2016xqb,SravanKumar:2018dlo} must be an effective version of a more fundamental one and 
the construction of our action \eqref{NAID} is a first step in achieving that.  } 

In the construction of \eqref{Genweyl}, we exploit the following {non-local} extension of identities of Weyl tensor \cite{Barvinsky:1994cg,Barvinsky:1994hw} 
\begin{equation}
	\begin{aligned}
		R^\sigma_\lambda  W_{\mu\nu\rho\sigma}W^{\mu\nu\rho\lambda} & = \frac{R}{4}W_{\mu\nu\rho\sigma}W^{\mu\nu\rho\sigma}  \\  \implies R^\sigma_\lambda\Oc\LF \square_s \RF W_{\mu\nu\rho\sigma}\Oc\LF \square_s \RF W^{\mu\nu\rho\lambda} & = \frac{R}{4}\Oc\LF \square_s \RF W_{\mu\nu\rho\sigma} \Oc\LF \square_s \RF W^{\mu\nu\rho\sigma}\, 
	\end{aligned}
\end{equation}
Following the papers \cite{Barvinsky:1994cg,Barvinsky:1994hw}, one can realize that at the cubic order in curvatures all possible terms can be recasted to have only d'Alembertians and not single covariant derivatives as follows\footnote{Indeed we can eliminate all terms with single covariant derivatives acting on curvature tensor by adding arbitrary number of total derivatives. For example, suppose we have a term of the form $\mathbb{R}\nabla_\mu\mathbb{R}\nabla^\mu \mathbb{R}$ in the action. Then we can eliminate this term by adding a total derivative of the form $\square \LF \mathbb{R}^3\RF$. We can continue this procedure for any arbitrary number of derivatives until we arrive at \eqref{cubpar}. }
\begin{equation}
	\Oc_1(\square_s)\mathbb{R}\Oc_2(\square_s)\mathbb{R}\Oc_3(\square_s)\mathbb{R}
	\label{cubpar}
\end{equation}
Then recalling that we use only terms involving Ricci scalar and Weyl tensor which do not spoil the Minkowski propagator, we arrive at \eqref{NAID}. It is useful to recall here the following identities of Weyl tensor which confirms that the cubic in Weyl tensor term that we have in \eqref{NAID} is indeed the only possible combination 
\begin{equation}
	\begin{aligned}
		& W_{[\mu\nu}^{\quad\gamma\lambda}\Oc\LF \square_s \RF W_{\gamma\lambda}^{\quad\alpha\beta}\Oc\LF \square_s \RF W_{\alpha]\sigma}^{\quad\mu\nu} = 0 \, ,\\ 
	&	W^{\mu\nu\rho\sigma}\Oc\LF \square_s\RF   W_{\mu\nu\rho\lambda}  = \frac{1}{4} \delta^\sigma_\lambda   W^{\mu\nu\rho\alpha}\Oc\LF \square_s \RF W_{\mu\nu\rho\alpha}\,, 
	\end{aligned}
	\label{weylid}
\end{equation} 
where in the first line we have complete anti-symmetrization over the five indices. 
Moreover, we can make the form-factors $\Cc\LF \square_s \RF$ to further depend on $R_s$ but we drop this generalization for brevity. 

It is important to mention that action (\ref{NAID}) can be viewed as leading terms in the series expansion in the approximation $R\ll M_P^2$, so one may expect other higher curvature terms to be added to build a full action.
{We note that the presence of higher derivative $R$ dependent terms with the Weyl tensor squared is essential for the theory to be ghost-free during inflation. } As a short comment, we can mention that even though a local term $\frac{RW^2}{M_p^2}$ renders the action non-renormalizable in a local theory of gravity, {one may expect that the presence of cubic curvature non-local terms together with the $R$-dependent $  \LF \frac{M_p^2}{2\Mc_s^2}+f_0\frac{R}{\Mc_s^2} \RF W_{\mu\nu\rho\sigma}\Fc_W\LF \square_s,\,R_s \RF W^{\mu\nu\rho\sigma}   $ retains the model to be renormalizable with expected additional conditions on the cubic non-local form factors in the 2nd, 3rd and 4th lines of \eqref{NAID}. }This, however, requires further study as this model is different from previously examined constructions like in \cite{Koshelev:2017tvv,SravanKumar:2018dlo}. Moreover, in the current manuscript, we focus on the study of inflation when it satisfies the condition $\bar{R}_{dS}\ll M_p^2$. Even though $\bar{R}_{dS}$ can be larger than $\Mc_s^2$, we should not worry at this stage about the validity of our model for all sub-Planckian energies. 

\subsection{Restrictions on form-factors (\ref{formfactl})}


The requirement that (\ref{mR2in}) is an exact background can be met by imposing the following constraints on the operators (see Appendix~\ref{App:EoM} for a detailed derivation using equations of motion following from the action \eqref{NAID})  
\begin{equation}
	\Fc_R\LF \frac{M^2}{\Mc_s^2} \RF = f_0=  \frac{M_p^2}{6M^2}\,,\quad \Fc^{\dagger}_R\LF \frac{M^2}{\Mc_s^2} \RF = 0\,,\quad \Lc_i\LF \frac{M^2}{\Mc_s^2} \RF=0\,,
	\label{condiq}
\end{equation}
where $^\dagger$ denotes derivative with respect to the argument. Since Weyl tensor vanishes on FLRW backgrounds, we do not get any constraints on the form-factors $\Fc_W\LF \square_s,\,\frac{2R}{3\Mc_s^2} \RF$, $\Cc_i\LF \square_s \RF$ and $\Dc_i\LF \square_s \RF$ as long as the background solution is concerned. Conditions on $\Fc_R$ are naturally the same as they were in the previous studies \cite{Koshelev:2016xqb,Koshelev:2017tvv} based on the action (\ref{AIDG}). The last equality in (\ref{condiq}) is a new condition that is sufficient, but not necessary though, and can be weakened as only 2 out of 3 operators $\Lc_i$ must vanish at the point $M^2/\Mc_s^2$ to have (\ref{mR2in}) as an exact background.\footnote{{The eigenvalue equation \eqref{mR2in} recently realized to give interesting cosmological solutions in a new class of non-local theories involving higher order scalar curvatures \cite{Dimitrijevic:2021aio}. }}


Another requirement is that the action \eqref{NAID} must be ghost-free containing only a scalaron and a massless graviton degree of freedom in the Minkowski and in the quasi-dS limits (i.e., $\bar{R}_{\rm dS}\approx \const$ and $f_0\bar{R}_{\rm dS}\gg M_p^2$, though still $\bar{R}_{\rm dS}\ll M_p^2$) during inflation. This leads to \cite{Koshelev:2016xqb,Koshelev:2017tvv}
\begin{equation}
	\begin{aligned}
		\Fc_R\LF \square_s \RF & = f_0 M^2 \frac{1-\LF 1-\frac{\square}{M^2}  \RF e^{\gamma_S\LF \square_s \RF} }{\square}\\
		\Fc_W\LF  \square_s,\, R_s\RF &  = 
		\frac{e^{\gamma_T
				\LF \square_s-\frac{2}{3} R_s \RF}-1}{\square_s-\frac{2}{3} R_s}\,\\
			\Dc_1\LF \frac{M^2}{\Mc_s^2} \RF &  =0\,,
	\end{aligned}
	\label{formfinal}
\end{equation}
where $\gamma$-s are entire functions. With the last condition in \eqref{formfinal}, we can notice that the contribution to the second order perturbed action from the 3rd line of \eqref{NAID} vanishes exactly (see Sec.~\ref{sec:pwnaid} for details). 
Additionally combining \eqref{condiq} with \eqref{formfinal}, one yields 
\begin{equation}
	\gamma_S\LF \frac{M^2}{\Mc_s^2} \RF =0 \implies \gamma_S\LF \square_s \RF = \LF \square_s-\frac{M^2}{\Mc_s^2} \RF \tilde\gamma_S\LF \square_s \RF \,. \label{entifR}
\end{equation}
where $\tilde{\gamma}_S$ is another entire function. 

Indeed, we count the degrees of freedom perturbatively by computing the second order action around a given background. Around Minkowski space-time, all the cubic and higher order curvature terms are irrelevant, and with \eqref{formfinal} the graviton propagator as a function of the 4-momentum square $p^2$ is given by \cite{Krasnikov:1987yj,Koshelev:2016xqb} 
\begin{equation}
	\Pi\LF p^2 \RF \sim -\frac{P^{(2)}}{p^2e^{\gamma_T\LF -p^2 \RF}} + \frac{P^{(0)}}{2p^2\LF 1+\frac{p^2}{M^2} \RF e^{\gamma_S\LF -p^2 \RF}}\,,
	\label{grprop}
\end{equation}
where $P^{(2)},\,P^{(0)}$ are spin projection operators. 

This implies the theory contains only a scalaron and a massless graviton in the spectrum \cite{Krasnikov:1987yj,Koshelev:2016xqb}. From the UV-completion point of view, it is expected that the propagator for the high momenta $p\to \infty$ is suppressed at least as in the Stelle gravity. Similarly, it is preliminary expected that $\Lc_i\LF \frac{p^2}{\Mc_s^2} \RF$, $\Cc_i\LF \frac{p^2}{\Mc_s^2} \RF$  and $\Dc_i\LF \frac{p^2}{\Mc_s^2} \RF$ at least do not grow in the high energy limit. Note that the action \eqref{NAID} with the form-factors \eqref{formfinal} is a generalization of the earlier studied non-local quadratic curvature gravity  \cite{Koshelev:2016xqb,Koshelev:2017tvv,Koshelev:2020foq} which can be seen as inclusion of the curvature dependence in the form-factors compared to the previous studies. This move is motivated by the understanding that the form-factors which depend on the d'Alembertian only cannot accommodate ghost-free conditions around Minkowski and (quasi)-dS spaces simultaneously. We note that the conditions (\ref{condiq}) are sufficient but not necessary. Indeed those conditions guarantee that (\ref{mR2in}) is an exact background solution as we demanded from the beginning of our consideration. However, regarding operators $\Lc_i$ it is enough that only any 2 out of 3 vanish at the point $M^2/\Mc_s^2$. Then any term in the equations of motion coming from $R^3$ part would vanish anyway. This can be traced by examining equations of motion given explicitly in Appendix~\ref{App:EoM}. However, imposing $\Lc_i\LF \frac{M^2}{\Mc_s^2} \RF =0$ on all 3 operators will make the contribution of cubic curvature term to the second order perturbed action vanish (see Sec.~\eqref{sec:pwnaid} for details) that will simplify our analysis.  We defer the other possible but complicated choices of $\Lc_i\LF \square_s \RF$ for future investigation, but we greatly expect that observational aspects of the model remain almost the same even in the general case.

\subsection{On the number of model parameters}
\label{sec:finitep}
Examining the construction of our action \eqref{NAID} which involve infinite order derivatives and infinite order terms in curvature, one might think the theory has infinite number of parameters. However we can end up with a theory with only finite parameter space. For example, let us take the entire functions  
\begin{equation}
	\Bigg\{\bar{\gamma}_S\LF \square_s \RF,\, \gamma_T\LF \square_s \RF \Bigg\}
	\label{entiref}
\end{equation}
defined in \eqref{formfinal} and \eqref{entifR} can be chosen to be either polynomials or functions of polynomials\footnote{For example, we can consider a choice prescribed in the context of super-renormalizable quadratic curvature non-local theories \cite{Koshelev:2016xqb} \begin{equation}
		\begin{aligned}
	\bar{\gamma}_S\LF \square_s \RF = \gamma_T\LF \square_s \RF & = 	\Gamma\Big(0,\,H_T(\square_s)^2\Big)+\gamma_E+\ln\Big(H_T(\square_s)^2\Big). 
	\end{aligned}
		\label{Tomfor}
	\end{equation}
	where $\Gamma\LF 0,\,z \RF = \int_z^\infty dt \frac{e^{-t}}{t}$ is the incomplete gamma function with first argument being zero, $\gamma_E$ is the Euler-Mascheroni constant and $H_T\LF \square_s \RF$ is some polynomial which determines the high energy behaviour of the graviton propagator around Minkowski. We can clearly see here that the entire functions depend on the choice of the polynomial function $H_T\LF \square_s \RF$, thus leading us to finite parameter space. {Worth to mention here that in \cite{SravanKumar:2018dlo} entire functions similar to \eqref{Tomfor} were phenomenologically studied with respect to the inflationary observables ($n_s,\,r,\,n_t$) in the quadratic curvature non-local theories.}  } which leads to finite parameter space and it can be observationally probed through inflationary observables that we shall study in the rest of the paper.  In this paper, we study two-point inflationary correlations which do not receive any contributions from the last three lines of the action in \eqref{NAID}. Thus, the parameter space associated with $\Lc_i\LF \square_s \RF,\, \Dc_i\LF \square_s \RF,\, \Cc_i\LF \square_s \RF$ is irrelevant for this paper, but we address it in the companion works \cite{Koshelev:2022bvg,KKS2}.

\subsection{Power spectrum predictions of generalized non-local $R^2$-like inflation with \eqref{NAID}} 
\label{sec:pwnaid}
In this section, we derive that the power spectrum predictions of $R^2$-like inflation in \eqref{NAID} using the several results obtained in \cite{Koshelev:2016xqb,Koshelev:2017tvv,Koshelev:2020foq} as explained below. Calculating the second order perturbation of the action \eqref{NAID} around the background solution $\bar{ \square} \bar{ R}=M^2 \bar R$, we obtain 
\begin{equation}
	\delta^{(2)}S_H^{\rm Non-local} =   \delta^{(2)}S_{R+R^2}^{\rm local}+\delta^{(2)}S_{R+R^2}^{\rm Non-local} + \delta^{(2)}S_{\mathbb{R}^3}^{\rm Non-local}
	\label{2rdv}
\end{equation}
where $S_{R+R^2}^{\rm local}$ is the local $R^2$ action and
\begin{equation}
	\begin{aligned}
		S_{R+R^2}^{\rm Non-local}  = &\, S_{R^2}^{\rm Non-local} + S_{W^2}^{\rm Non-local} \\  =
		&\, \frac{1}{2}\int d^4x\sqrt{-g} \Bigg\{R \Bigg[\Fc_R\LF\square_s\RF-f_0\Bigg]R\\ &+\LF\frac{M_p^2}{2\Mc_s^2}+ f_0\frac{R}{\Mc_s^2} \RF  W_{\mu\nu\rho\sigma}{\mathcal{F}}_{W}\left(\square_{s},\,\frac{R}{\Mc_{s}^2}\right)W^{\mu\nu\rho\sigma}\Bigg\} 
		\label{nlcl2q}
	\end{aligned}
\end{equation}
and $S_{\mathbb{R}^3}^{\rm Non-local}$ constitute from the last 3-lines in \eqref{NAID} which are cubic in curvature.
Remarkably, applying $\Lc_i\LF \frac{M^2}{\Mc_s^2}\RF =0 $ from \eqref{condiq}, we can see that the cubic non-local term has no contribution to the above second order action around the flat FLRW background satisfying\footnote{
	It is also not very surprising that the cubic scalar curvature term in the action \eqref{NAID} do not affect the second order action around the inflationary solution  $\bar{\square}\bar{R}=M^2\bar{R}$. An analogous simple example is the following action 
	\begin{equation}
		S^{\rm Min}_{R^3} = \frac{1}{2}\int d^4x\sqrt{-g}\,  \LT M_p^2R+ f_0R^2+f_0\Lambda^2R^3 \RT\,
	\end{equation}
	where $\Lambda$ is some mass scale. The above Lagrangian has the following property that the cubic term $R^3$ does not effect the second order action (or the linearized equations of motion) around Minknowski, however, it is obviously not true in other backgrounds. If we study inflation in any local higher curvature extension of $R^2$, we do modify inflationary solution unless we assume $R^2$ is the most relevant term during inflation and remaining terms are highly negligible. In fact, the general statement is that slow-roll inflation in local $f(R)$ gravity occurs for the range of $R$ for which the Lagrangian density ${\cal L}$ is close to $R^2$, namely, ${\cal L}=A(R)R^2$ where $A(R)$ is a slowly changing function of $R,~|\frac{d\ln A}{d\ln R}|\ll 1$~\cite{Appleby:2009uf}. In our case the cubic scalar curvature non-local terms \eqref{NAID} are introduced in order not to effect the second order scalar perturbations around the inflationary phase given by $\bar{\square} \bar{ R} = M^2\bar{R}$. } \eqref{mR2in}. 
\begin{equation}
	\delta^{(2)}S_{\mathbb{R}^3} \Bigg\vert_{\bar{\square} \bar{ R} = M^2\bar{R}} = 0\,.
\end{equation}

Therefore, non-local contribution to the two-point inflationary correlations comes only from the second variation of \eqref{nlcl2q}. The cubic in curvature terms in \eqref{NAID} will give non-trivial shapes of non-Gaussianities which will be investigated in a separate paper \cite{Koshelev:2022bvg}.  
We present in detail the computation of the second order action \eqref{2rdv} and analyze the corresponding perturbed degrees of freedom in the quasi-dS limit in Appendix~\ref{App:pEoM}. After long computations, we deduce that our second order action in \eqref{2rdv} exactly coincides with the result obtained in (4.5) of \cite{Koshelev:2017tvv}. As a result, the predictions for scalar and tensor power spectra become exactly the same as in the theory studied in  \cite{Koshelev:2016xqb,Koshelev:2017tvv,Koshelev:2020foq}. 
The crucial difference in our study is that we consider a choice of $\Fc_R\LF \square_s \RF,\,\Fc_W\LF \square_s \RF$ that does not explicitly depend on the background curvatures during inflation which was the case of \cite{Craps:2014wga,Koshelev:2017tvv,Koshelev:2020foq}. Therefore, the scalar power spectrum $\Pc_\Rc$ and spectral index $n_s$ of generalized non-local $R^2$ inflation are 
\begin{equation}
	\Pc_{\Rc} (k)\approx \frac{1}{3f_0\bar{R}_{\rm dS}} \frac{H^2}{16\pi^2\epsilon^2}\Bigg\vert_{k=aH}\,,\quad n_s-1 \equiv \frac{d\ln \Pc_{\Rc}}{d\ln k }\Bigg\vert_{k=aH} \approx -\frac{2}{N}\,, 
	\label{pr}
\end{equation}
where $a(t)H(t)$ is estimated at the first Hubble radius crossing during inflation for a given mode wavenumber $k$. 
The above result \eqref{pr} is almost independent of $\Fc_R\LF \square_s\RF$ as far as $M^2\ll \Mc_s^2$, and the dimensionless coefficients of $\Fc_R\LF \square_s\RF$ \eqref{formfinal} are of the $O(1)$ (see \cite{Koshelev:2017tvv,Koshelev:2020foq} for further details, where it was shown that the non-local corrections to $n_s$ are of the order of $O\LF\frac{M^4}{\Mc_s^4}\RF$). 

\begin{figure}[h!]
	\centering	\includegraphics[width=2.8in]{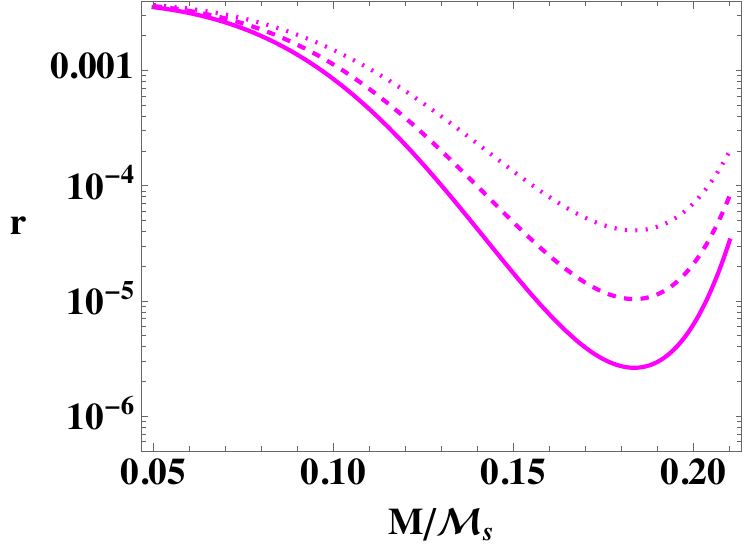}\quad \includegraphics[width=2.8in]{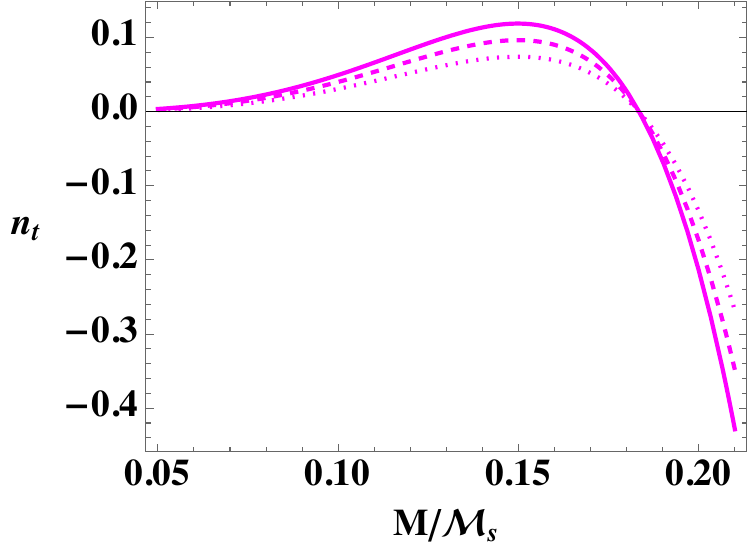}\quad \includegraphics[width=2.8in]{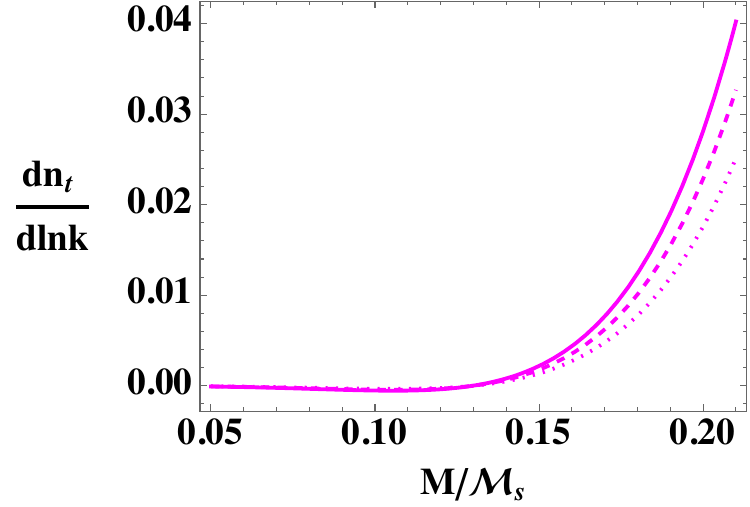}\quad \includegraphics[width=2.8in]{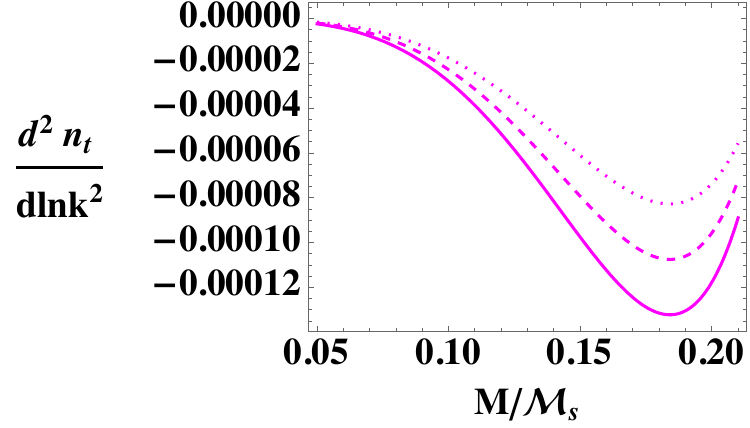}
	\caption{In the upper left and right panel we plot the tensor-to-scalar ratio ($r$) and the tensor tilt ($n_t$) versus the ratio of scalaron mass $(M)$ and the scale of non-locality ($\Mc_s$) respectively. In the lower panel we plot the running $\LF\frac{dn_t}{d\ln k}\RF$ and running of running of the tensor spectral index  $\LF\frac{d^2n_t}{d\ln k^2}\RF$ versus $\frac{M}{\Mc_s}$. The parameters for all these plots are $\LF \beta_1,\,\beta_2 \RF = \LF 0.8,\, 1.215 \RF $,\,$\LF \beta_1,\,\beta_2 \RF = \LF 0.65,\, 0.117 \RF $,\,$\LF \beta_1,\,\beta_2 \RF = \LF 0.5,\, 0.09 \RF $ for full, dashed and dotted lines in magenta respectively. }
	\label{RS1}
\end{figure}

\begin{figure}[h!]
	\centering
	\includegraphics[width=2.8in]{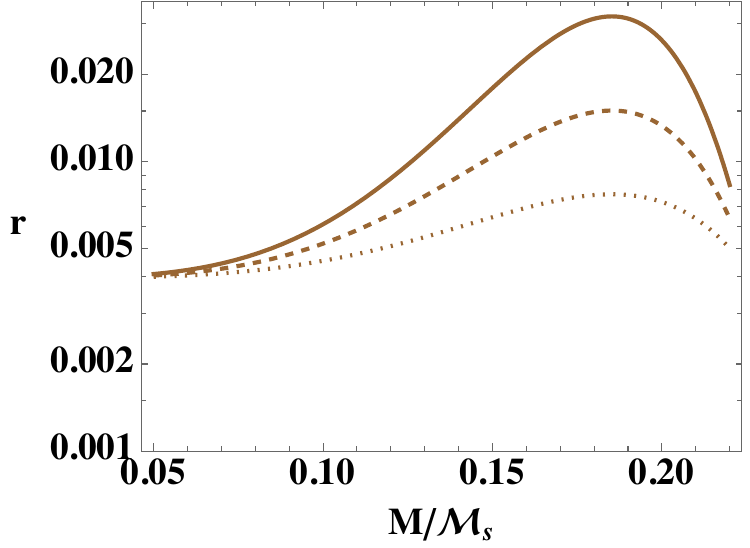}\quad \includegraphics[width=2.8in]{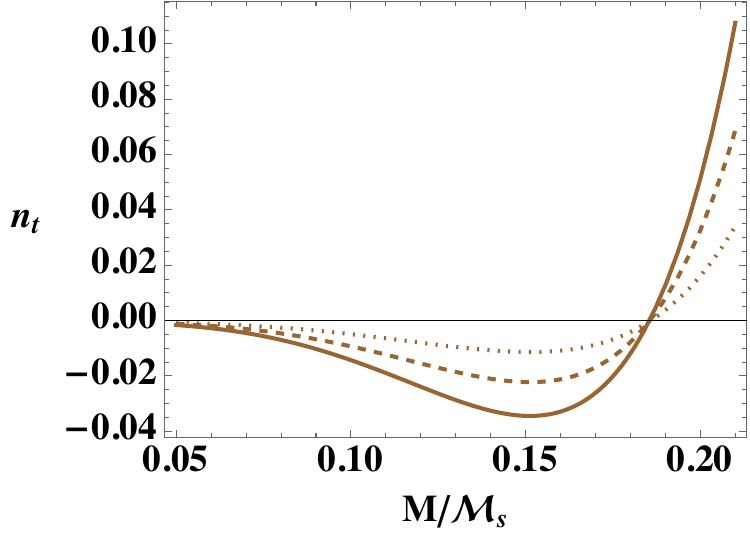}\quad \includegraphics[width=2.8in]{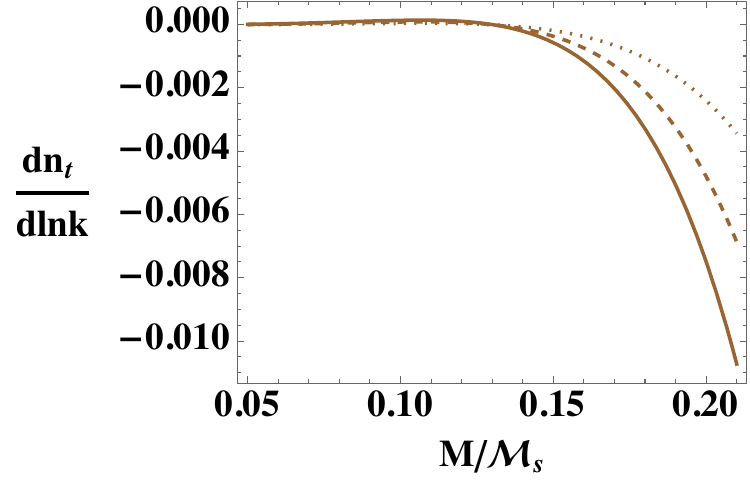}\quad \includegraphics[width=2.8in]{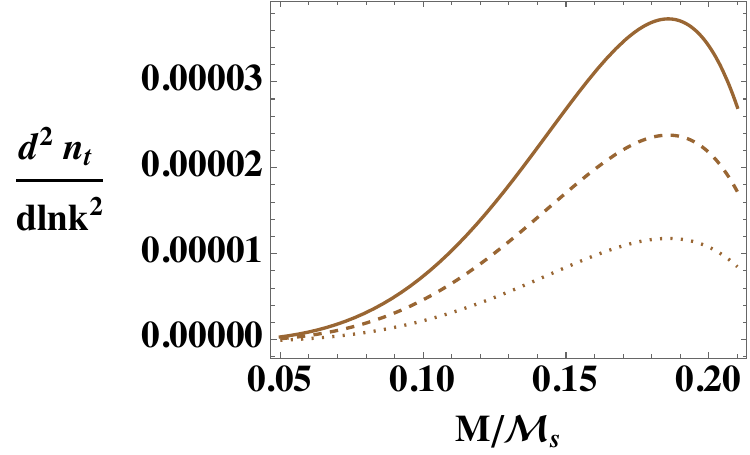}
	\caption{In the upper left and right panel we plot the tensor-to-scalar ratio ($r$) and the tensor tilt ($n_t$) versus the ratio of scalaron mass $(M)$ and the scale of non-locality ($\Mc_s$) respectively. In the lower panel we plot the running $\LF\frac{dn_t}{d\ln k}\RF$ and running of running of the tensor spectral index  $\LF\frac{d^2n_t}{d\ln k^2}\RF$ versus $\frac{M}{\Mc_s}$. The parameters for all these plots are $\LF \beta_1,\,\beta_2 \RF = \LF 0.625,\, 0.1125 \RF $,\,$\LF \beta_1,\,\beta_2 \RF = \LF 0.4,\, 0.072 \RF $,\,$\LF \beta_1,\,\beta_2 \RF = \LF 0.2,\, 0.036 \RF $ for full, dashed and dotted lines in brown respectively.}
	\label{RS2}
\end{figure}

The inflationary tensor power spectrum of generalized non-local $R^2$ model
\begin{equation}
	\begin{aligned}
		\Pc_T & = \frac{1}{12\pi^2f_0} \LF 1-3\epsilon \RF e^{-2\gamma_T\LF \frac{-\bar{	R}_{dS}}{6\Mc_{s}^2} \RF} \Bigg\vert_{k=a H}\,,
	\end{aligned}
	\label{pt}
\end{equation}
Computing the tensor-to-scalar ratio from \eqref{pr} and \eqref{pt} we obtain 
\begin{equation}
	r = \frac{12}{N^2} e^{-2\gamma_T\LF -\frac{\bar{	R}}{6\Mc_s^2} \RF}\Bigg\vert_{k=aH}\,. 
	\label{t2s}
\end{equation}
The tensor spectral index and its running and running of running can be computed as
\begin{equation}
	\begin{aligned}
			n_t \equiv \frac{d\ln\Pc_T}{d\ln k }\Bigg\vert_{k=aH} & \approx -\frac{3}{2N^2}-\LF \frac{2}{N}+\frac{1}{N^2} \RF \frac{\bar{	R}_{\rm dS}}{6\Mc_s^2} \gamma_T^{\dagger}\LF -\frac{\bar{	R}_{\rm dS}}{6\Mc_{s}^2} \RF \, \\ 
		\frac{dn_t}{d\ln k} \Bigg\vert_{k=aH} & \approx -\frac{3}{N^3}-\frac{1}{N^3}\frac{\bar{R}_{\rm dS}}{6\Mc_s^2}\gamma_T^{\dagger}\LF -\frac{\bar{	R}_{\rm dS}}{6\Mc_{s}^2} \RF -\frac{1}{18N^2}\frac{\bar{	R}_{\rm dS}^2}{\Mc_{s}^4}  \gamma_T^{\dagger\dagger}\LF -\frac{\bar{	R}_{\rm dS}}{6\Mc_{s}^2} \RF \,\\
		\frac{d^2n_t}{d\ln k^2} \Bigg\vert_{k=aH} & \approx -\frac{9}{N^4}-\frac{1}{3N^4}\frac{\bar{R}_{\rm dS}}{\Mc_s^2}\gamma_T^{\dagger}\LF -\frac{\bar{R}_{\rm dS}}{6\Mc_s^2} \RF -\frac{1}{12N^3}\frac{\bar{R}_{\rm dS}^2}{\Mc_s^4}\gamma_T^{\dagger\dagger}\LF -\frac{\bar{R}_{\rm dS}}{6\Mc_s^2}\RF \\ &\quad -\frac{1}{108N^3}\frac{\bar{R}_{\rm dS}^3}{\Mc_s^6}\gamma_T^{\dagger\dagger\dagger}\LF -\frac{\bar{R}_{\rm dS}}{6\Mc_s^2}\RF
	\end{aligned}
	\label{ntt}
\end{equation}
where $^{\dagger},\,^{\dagger\dagger},\,^{\dagger\dagger\dagger}$ indicate first, second and third derivative with respect to the argument, we used $\frac{d}{d\ln k}\approx -\LF 1+\frac{1}{2N} \RF\frac{d}{dN}$ and $\frac{d\bar{R}_{\rm dS}}{dN}\approx 2\bar{R}_{\rm dS}\epsilon$. 
From \eqref{pt}, \eqref{t2s} and \eqref{ntt} we can conclude the following 
\begin{itemize}
	\item The tensor power spectrum is modified due to the non-local contributions from the higher curvature non-local terms involving Ricci scalar and Weyl tensor in \eqref{NAID} i.e., due to the effect of \eqref{Genweyl} in the limit $M_p^2 \ll f_0 R$.
	\item In comparison with local $R^2$ gravity, the tensor power spectrum contains a strong scale dependence due to the exponential term $e^{\gamma_T\LF\frac{\bar{R}_{\rm dS}}{\Mc_s^2}\RF}$ where $\bar{R}_{\rm dS}(k)$ depends on wave number $k$. 
	\item With tensor tilt and its running and running of running computed in from \eqref{ntt} we can probe $\gamma_T\LF\frac{\bar{R}_{\rm dS}}{\Mc_s^2}\RF$ and reconstruct the form-factor \eqref{formfinal}  from the future primordial gravitational wave observations \cite{CMB-S4:2020lpa,Calcagni:2020tvw}. 
	\item The single field tensor consistency relation $r=-8n_t$ gets violated solely due to the modification of tensor-power spectrum. 
\end{itemize}
To illustrate the above conclusions we consider a choice of entire function
\begin{equation}
	\gamma_T \LF  \square_s-\frac{2R}{3\Mc_{s}^2}  \RF =  \beta_1\LF \square_{s}- \frac{2R}{3\Mc_{s}^2} \RF^2+\beta_2\LF\square_{s}- \frac{2R}{3\Mc_{s}^2}\RF^3 \,. 
	\label{fw1}
\end{equation} 
and present the predictions of the model for the different ranges of parameter space $\LF \beta_1,\,\beta_2 \RF$ in Fig.~\ref{RS1} and Fig.~\ref{RS2}. 

From \eqref{t2s} and \eqref{ntt} we can derive the following relation between tensor tilt $n_t$, running of tensor to scalar ratio $\frac{d\ln r}{d\ln k}$. 
\begin{equation}
	n_t \approx -\LF 1-n_s \RF + \frac{d\ln r}{d\ln k}
	\label{newcon}
\end{equation}

\textbf{Nature of scalaron:}
From \eqref{s2s} we can extract that the sound speed of curvature perturbation is unity.  Therefore, we have a canonical scalar field propagating during inflation similar to the local $R^2$ theory. Thus, we can also conclude that there is no super-horizon evolution of the curvature perturbation $\Rc$ and it can be assumed to be approximately constant on superhorizon scales as its solution generally contain a growing mode and a decaying mode \cite{DeFelice:2010aj}
\begin{equation}
	\Rc = c_1 +c_2\int \frac{dt}{a^3\epsilon}\,. 
\end{equation}

\section{Non-local $R^2$-like inflation versus effective field theory of scalar field inflation}
\label{sec:eftcom}

The goal of this section is to provide a qualitative and somewhat quantitative understanding of how the framework of inflation in non-local gravity \eqref{NAID} is different from the effective field theory  of (single-field) inflation (EFTI) \cite{Cheung:2007st,Weinberg:2008hq}. Of course, the obvious difference between the two is \say{locality}, but here we discuss in detail how EFTI cannot capture everything about our ignorance of UV-completion. We argue that even though EFTI is a useful tool to probe the unknown UV-physics, the well-motivated inputs from UV-completion point of gravity may shed new lights on inflationary cosmology beyond what is promised by EFTI.

\subsection{Nature of primordial fields in EFTI}
Ever since the EFTI was proposed  \cite{Cheung:2007st,Weinberg:2008hq,Creminelli:2014wna,Creminelli:2015oda}, it is widely understood and established that EFTI captures all our ignorance about UV-completion and the approach unifies all our models of inflation through the so-called EFT parameters. As far as the inflationary two-point correlations are concerned, EFTI prescribes that everything about inflation can be captured by the following set of slowly varying time-dependent parameters 
\begin{equation}
	\Biggl\{H_{\rm \inf},\,\epsilon_E,\,\eta_E,\,c_s,\,c_t\Biggr\}
\end{equation}
where $H_{\rm \inf}$ is the nearly constant Hubble parameter during inflation in an EFT.  The corresponding slow-roll parameters are 
\begin{equation}
	\epsilon_E = -\frac{\dot{H}_{\rm inf}}{H_{\rm inf}^2},\quad \eta_E =  \frac{\dot{\epsilon}_E}{H_{\inf}\epsilon_E}\,. 
\end{equation}
The additional two-parameters (which are also nearly constant) $\LF c_s,\,c_t \RF$ are the sound speed of scalar and tensor perturbations defined through their second order actions 
\begin{equation}
	\begin{aligned}
		\delta_s^{(2)}S_{\rm EFTI} & = \int d^4x \sqrt{-g}\Bigg[ \frac{M_p^2}{c_s^2}\epsilon\LF \dot{\zeta}^2 -c_s^2\frac{\LF\pd_i\zeta\RF^2}{a^2}\RF \Bigg]\\
		\delta_t^{(2)}S_{\rm EFTI} & = \int d^4x \sqrt{-g}\Bigg[ \frac{M_p^2}{c_t^2}\epsilon\LF \dot{h}_{jk}^2 -c_t^2\frac{\LF\pd_i h_{jk}\RF^2}{a^2}\RF \Bigg]
	\end{aligned} 
	\label{EFTsoa}
\end{equation} 
where $\zeta=-{\cal R}$ is the curvature perturbation defined in the so-called Unitary gauge of EFTI and $h_{ij}$ is the tensor standard tensor perturbation. During inflation the scalar perturbation can be approximated to be massless in the first approximation. In standard canonical scalar field inflation (for example Starobinsky or Higgs inflation \cite{Kehagias:2013mya}), the sound speeds $c_s=c_t =1$ which means that the speed of propagation of scalar and tensor perturbed modes equals to the velocity of light.

The construction of EFTI conveys that our ignorance of UV-completion beyond standard canonical scalar field inflation should be encoded in $\LF c_s,\,c_t \RF$ different from Unity.   As a consequence, the key observables of inflation are modified in the leading order of the slow-roll approximation as \cite{Burgess:2017ytm,Baumann:2015xxa}
\begin{equation}
	n_s = 1-2\epsilon_E-\eta_E-\epsilon_s,\quad r= 16\frac{c_s}{c_t}\epsilon_E,\quad n_t= -2\epsilon_E -\epsilon_t,\quad r\neq -8n_t\,.
	\label{nsrntEFTI}
\end{equation}
where $\epsilon_s=\frac{\dot{c}_s}{Hc_s}\ll 1,\, \epsilon_t = \frac{\dot{c}_t}{Hc_t}\ll 1$ are the slow-roll parameters associated with the sound speeds. Therefore, the central observationally relevant message from EFTI is that any violation of single field tensor consistency relation $r = -8n_t$ must arise from $0\leq \LF c_s,\,c_t \RF \leq 1$. The generalized non-local $R^2$-like inflation is a counterexample to this which we discuss below. 

\subsection{Differences between non-local $R^2$-like inflation and EFTI}
Here we make theoretical and observational comparisons between our framework of geometric inflation with \eqref{NAID} and the EFTI. In the theoretical part we exploit the differences between the construction of EFTI and  \eqref{NAID} with respect to the nature of perturbed degrees of freedom and the cut-off scale and the roads to UV-completion. In the observational  part we make qualitative and few quantitative remarks on distinguishing EFTI with our generalized non-local $R^2$-like inflation. 
\subsubsection{With respect to theoretical framework} 
Here we provide a comparative discussion of EFTI \cite{Cheung:2007st,Weinberg:2008hq} and our non-local gravity inflation. 
\begin{itemize}
	\item EFTI is motivated from particle physics and identifying carefully the symmetries based on the adiabatic evolution of the inflationary background. EFTI construction by Cheung et al \cite{Cheung:2007st} captures the nature of inflationary perturbations and their interactions around a standard slow-roll inflation by exploiting time dependent spatial diffeomorphisim invariance in the frame of the broken time diffeomorphisms. This procedure further employs the so-called Stuckelberg trick to non-linearly recover time-diffeomorphism invariance and to obtain an effective Lagrangian for a scalar field that is valid below a cut-off scale $\Lambda \gtrsim H_{\rm inf}$. Our geometric framework of inflation studied in this paper goes beyond any of the assumptions of symmetries and perturbative expansion of terms of order $\frac{H}{\Lambda}$ in the EFTI. We collected all possible curvature invariant terms one can write around an exact inflationary background \eqref{mR2in}. We have ended up with infinite number of curvature invariant terms in the action \eqref{NAID} which cannot be truncated at any order as far as the $\Mc_s^2\lesssim \bar{	R}_{\rm dS}\ll M_p^2$. Moreover, our diffeomorphism invariant action \eqref{NAID} captures a significant set of terms which can be relevant for quantum gravity. 
	\item The construction of EFTI relies on an existence of a cut-off scale $\Lambda>H_{\inf}$, and it assumes that all effects (of possible existence of heavy modes that can be integrated out) above this cut-off scale can be encoded in terms of EFTI parameters.  In our construction, we do have a scale $\Mc_s \gtrsim H$ which is the scale of non-locality, but we have found that collecting all possible non-local terms we get new effects for inflationary correlations which cannot be mapped to EFTI parameters. Indeed comparing \eqref{EFTsoa} with \eqref{s2s} and \eqref{s2h}, we can deduce that the sound speeds of perturbations in the generalized non-local $R^2$-like inflation are Unity. But in our study we do obtain violation of tensor consistency relation (read from \eqref{t2s} and \eqref{ntt}) due to the non-local effect, more precisely due to the terms involving Weyl tensor in \eqref{NAID}. 
	Therefore,  perturbations in generalized non-local $R^2$-like inflation do not belong to any special case of EFTI which thus proves that the assumptions of EFTI capturing all possible UV-signatures breaks down at least in our context. 
	\item In EFTI the sound speeds deviation from Unity is signature of non-relativistic effects that can occur in the low-energy (at the scale of inflation) theory which may or may not persist in the UV-theory. For example, if the fundamental theory is Lorentz invariant, the non-relativistic effects are emergent in the low-energy limit. In the alternative case, Lorentz invariance may not be the fundamental symmetry in the UV \cite{Baumann:2015nta,Aoki:2021ffc}. Our theory \eqref{NAID} is Lorentz invariant and we do not have any non-relativistic regime at any energy scale. Moreover, we entirely fix the gravitational degrees of freedom at all energy scales (see Sec.~\ref{sec:pwnaid} and Appendix~\ref{App:pEoM}).
	\item In \cite{Weinberg:2008hq}  Weinberg has proposed a version of EFTI by constructing a covariant action which is a bit more general than the one by Cheung et al \cite{Cheung:2007st}. Weinberg has introduced new terms that indeed are relevant in the EFTI,  especially the term relevant to our discussion is\footnote{In the Weinberg's paper \cite{Weinberg:2008hq} the scalar field $\phi$ is dimensionless and the scalar field kinetic term is written as $ -\frac{\Lambda^2}{2} \LF\pd\phi\RF^2$, but here we do the rescaling $\phi\to \frac{\phi}{\Lambda}$ that is reflected in \eqref{Weyl-Weinberg}.}
	\begin{equation}
		\int d^4x\sqrt{-g}\Biggl[ f_1\LF \frac{\phi }{\Lambda}\RF W^{\mu\nu\rho\sigma} W_{\mu\nu\rho\sigma} \Biggr]
		\label{Weyl-Weinberg}
	\end{equation} 
	where $\phi$ is a canonically normalized inflaton field or the Goldstone boson in the language of EFTI in  \cite{Cheung:2007st}. The presence of Weyl square terms in the action does not effect the background equations for FLRW, but they do effect perturbations. However, for any effective scale $\Lambda < M_p$,  Weyl square term \eqref{Weyl-Weinberg} may not be the lowest order term and in principle we could add higher derivative and higher curvature terms which do not effect the FLRW solutions  
	\begin{equation}
		\begin{aligned}
		\int d^4x\sqrt{-g}\Biggl[& f_1\LF \frac{\phi }{\Lambda}\RF  W^{\mu\nu\rho\sigma}\Fc_1\LF \frac{\square}{\Lambda^2},\,\frac{R}{\Lambda^2}\RF W_{\mu\nu\rho\sigma} \Biggr]
		\label{Weyl-Weinberg1}
		\end{aligned}
	\end{equation} 
	where $ \Fc_1\LF \frac{\square}{\Lambda^2},\,\frac{R}{\Lambda^2} \RF$ is an analytic non-polynomial functions of d'Alembertian and the Ricci scalar. The Weyl tensor contributions in \eqref{Weyl-Weinberg1} cannot be truncated to any finite order especially if $R\gtrsim \Lambda^2,\,\square\gtrsim \Lambda^2$ and if $\phi\sim O\LF M_p\RF$ during inflation\footnote{This is true in the context of large field inflation, e.g., local $R^2$ or Higgs inflation.} one can expect $f_1\LF \frac{\phi}{\Lambda} \RF \gg 1$. The scalar field model of inflation with \eqref{Weyl-Weinberg1} is not the subject of present paper at all, but \eqref{Weyl-Weinberg1} expands the current understanding of EFTI and inevitability of the higher curvature and higher derivative terms, especially when the square of the cut-off scale is comparable to Ricci scalar. Analogously, we can think of \eqref{Weyl-Weinberg} as the one similar to the term  $\frac{R}{\Mc_s^2}W^{\mu\nu\rho\sigma}W_{\mu\nu\rho\sigma}$ in our action \eqref{NAID}, and it is indeed not the lowest order term of the action. 
	Therefore, our construction in a way goes beyond the EFTI proposed by Weinberg \cite{Weinberg:2008hq}.  Moreover, including just the Weyl square terms like \eqref{Weyl-Weinberg} can lead to non-trivial sound speed for tensor modes as well as for a tensor ghost \cite{Baumann:2015xxa}, but it need not be the case if we sum up all possible  terms (which is precisely true in our case). 
	Therefore, in this sense EFTI cannot actually capture fully the UV-complete theory, but on the contrary, EFTI could lead us to the physics that may not be compatible with UV-theory. 
	This is certainly because EFTI is built on just writing the lowest order terms with symmetry assumptions which are subjected to only particular class of UV-complete theories \cite{Burgess:2017ytm}.
	\item Since our action \eqref{NAID} contains higher derivative terms involving Weyl tensor, it is not possible to rewrite this action in the Einstein frame by using a conformal transformation of the metric. This means that we cannot cast our action \eqref{NAID} to the form where gravity part is described by the Einstein-Hilbert term only and all the rest is written as a (non-local) scalar field with a potential. This can be also understood focusing on the fact that the graviton propagator \eqref{grprop} in our case gets modified compared to GR exactly due to the presence of the Weyl tensor terms. This is a significant difference in comparison to many models of scalar-tensor theories that belong to the class of EFTI \cite{Burgess:2017ytm} where inflation can be rewritten in the Einstein frame.
\end{itemize}

\subsubsection{With respect to inflationary observables}
Here we discuss the ways to distinguish EFTI with non-local gravity inflation in the context of inflationary observables built from two-point correlation functions. With the 3-point correlations (i.e., Non-Gaussianities) which we study in the companion paper \cite{Koshelev:2022bvg}, we can further distinguish the non-local gravity with EFTI.
\begin{itemize}
	\item Inflationary observables \eqref{inobservables} in the EFTI depends on the time dependent parameters $\LF c_s,\,c_t \RF$ which are determined by the background scalar field evolution whereas the in the case of non-local gravity we have constant parameters associated with the choice of entire functions \eqref{entiref}. 
	\item Since the sound speeds must be bounded $0\leq \LF c_s,\,c_t \RF \leq 1$ to avoid instabilities, they must be slowly-varying or almost constant during inflation. 
	Let us consider EFTI with $c_t=1$ and $c_s\neq 1$ i.e., the case of a non-canonical scalar field inflation in which $r= -8c_sn_t$, $n_t = -2\epsilon$ and $r < -8n_t$. In our non-local gravity inflation  tensor tilt is modified as well as its running \eqref{ntt}.  We can see this in Fig.~\ref{RS1} and Fig.~\ref{RS2}. Moreover, we can have $r> -8n_t$ and also blue tensor tilt $n_t$ that is not possible to achieve in the context of inflation with non-canonical scalar field \cite{Chen:2006nt}. Inflationary models with $c_t\neq 1$ suggested within EFTI  (and also generalized Galileon theories \cite{Kobayashi:2011nu} and the models where inflaton is coupled to Weyl tensor square term like in \eqref{Weyl-Weinberg} \cite{Baumann:2015xxa} ) can lead to varied predictions with respect to $n_t$ and its running, which are in common with our inflation in non-local gravity. But $c_t$ is expected to be slowly varying, so we greatly expect $\frac{dn_t}{d\ln k} \ll n_t$ in these models. But in our case we can read from Fig.~\ref{RS1} and Fig.~\ref{RS2} that both tensor tilt and its running can be of comparable order $\frac{dn_t}{d\ln k} \sim  O \LF n_t \RF$, or running of $n_t$ can dominate over $n_t$, that is $\frac{dn_t}{d\ln k} \gg n_t $ and vice versa. These new features place the non-local gravity inflation as the unique interesting target in the scope of future observations \cite{CORE:2016ymi}. 
	\item Moreover, possible relation between $c_s$ and $c_t$ via (dis-) conformal transformation sheds further light on the inflationary observables  \cite{Creminelli:2014wna,Burrage:2016myt} which might help us to distinguish the EFTI and our framework of non-local gravity where we cannot perform a transformation to the Einstein frame. 
\end{itemize}

Primordial non-Gaussianities is another powerful window to further probe the signatures of generalized non-local $R^2$-like inflation which we shall be presenting in the companion paper \cite{Koshelev:2022bvg}. In this section we have only discussed EFTI of single field inflation in the face of non-local gravity inflation, but there exist two other categories known as multi-field inflation (where more than one scalar fields of comparable mass drive inflation) and the quasi-single field inflation(where inflation is driven by a single scalar field with additional degrees of freedom whose masses are being of the order of Hubble perarameter square \cite{Chen:2015lza}).  These scenarios lead to notable observational signatures in the context of primordial non-Gaussianities \cite{Cabass:2022avo}. Therefore, we defer the detailed discussion of these effects to the paper on primordial non-Gaussianities in non-local gravity inflation \cite{Koshelev:2022bvg}. 

\section{Conclusions and outlook}

In this paper, we take a significant step forward on understanding cosmic inflation beyond the $R^2$ model. Our quest for a most general theory of gravity consistent with $R^2$-like inflation has led to the construction of gravity theory containing terms of infinite order in derivatives and curvature which goes beyond the previous construction of non-local gravity theories \cite{Koshelev:2020xby}. Our theory opens doors for a new understanding of quantum gravity. So far only non-local quadratic curvature theories were expected to be good candidates for quantum gravity, but here we propose higher order terms which are inevitable if we wish to explain inflation in non-local gravity. 

{The quadratic curvature non-local theories \eqref{AIDG}~\cite{Tomboulis:1997gg,Modesto:2011kw,Biswas:2011ar,Koshelev:2016xqb} are so far shown to be super-renormalizable only by power counting methods around the Minkowski background.\footnote{{Renormalizability around de Sitter is studied in a class of non-local theories where $f_0=0~$\cite{Koshelev:2017bxd}.  But assuming $f_0=0$, one cannot get $R^2$-like inflation in non-local gravity~\cite{SravanKumar:2018dlo}.}} Although super-renormalizability around Minkowski is a promising aspect, it is greatly expected that one must go beyond \eqref{AIDG} for a consistent formulation of quantum gravity. In this regard, our formulation of \eqref{NAID} as an extension of \eqref{AIDG} valid up to the inflationary energy scale is a pertinent advancement towards a full theory of quantum gravity. In our opinion, the presence of non-local term $RW_{\mu\nu\rho\sigma}\Fc_W\LF \square_s,\, R_s \RF W^{\mu\nu\rho\sigma}$ does not necessarily spoil the power counting renormalizability {because of the additional presence of cubic and higher order curvature non-local terms}. {Indeed if one considers the generalized ghost-free form-factor \eqref{formfinal} of the action \eqref{NAID}, the N-point graviton vertices arising from $W_{\mu\nu\rho\sigma}\Oc\LF \square_s \RF W^{\mu\nu\rho\sigma}$ dominate over the $N-$point vertices arising from $R^{N-2}W_{\mu\nu\rho\sigma}W^{\mu\nu\rho\sigma}$ term in the high energy limit. {Furthermore, one must also notice that we have similar cubic curvature terms with different structure of non-locality i.e., $f_0R_sW_{\mu\nu\rho\sigma}\Fc\LF  \square_s,\,R_s \RF\Big\vert_{R_s=0} W^{\mu\nu\rho\sigma}$ and $\frac{f_0\lambda_R}{\Mc_s^2}\Dc_1\LF \square_s \RF R\Dc_2\LF \square_s\RF W_{\mu\nu\rho\sigma}\Dc_3\LF  \square_s \RF W^{\mu\nu\rho\sigma}$ whose vertices can cancel with each other with appropriate conditions on the different form factors. }
			Therefore, it is very unlikely that the Ricci scalar-dependent terms spoil the traditional power counting renormalizability around Minkowski background which is studied in~\cite{Koshelev:2016xqb}. Despite this observation, we would like to point out that renormalizability might further require more additional terms to those present in \eqref{NAID}.  }  {This detailed analysis of the renormalizability of the action \eqref{NAID} and further possible extension of it} is beyond the scope of this paper, so we defer its detailed investigation for the future. }
{In this paper, w}e constrained the form-factors in the action \eqref{NAID} with the aim of achieving ghost-free degrees of freedom in the Minkowski and quasi-dS limits, but it is important to constrain them from the UV-completion point of view. In this regard, studying scattering amplitudes and perturbative renormalizability of \eqref{NAID} is the most important direction to progress further. In Appendix~\ref{App:pEoM} we discuss how the form-factors can lead to infinitely many complex conjugate poles in the propagator of perturbed modes in curved space-time. In our study, we assumed zero initial conditions for degrees of freedom associated with these complex conjugate poles, but it is worth investigating this issue further by formulating the optical theorem in curved space-time rigorously. 

In our formulation of gravity action \eqref{NAID}, we explicitly dropped terms involving Ricci tensor, but it is interesting to study further the implications of including these terms for quantum gravity and inflationary observables. Also in this study, we do not include curvature terms that violate parity, but in principle, one can include them and compute their implications in the context of inflationary observables. Finally, it is interesting to explore the frameworks of multi-field and quasi-single field inflation in the non-local context that might lead us to new directions in the field.

\label{sec:Conc}

\acknowledgments
AK is supported by FCT Portugal investigator project IF/01607/2015. This research work was supported by grants  UID/MAT/00212/2019, COST Action CA15117 (CANTATA). KSK acknowledges the support from JSPS and KAKENHI Grant-in-Aid for Scientific Research No. JP20F20320 and No. JP21H00069. KSK would like to thank the Royal society for the Newton International Fellowship. AAS was supported by the RSF grant 21-12-00130.  We thank L. Buoninfante, A. De Felice, T. Noumi, A. Tokareva and especially M. Yamaguchi for very useful discussions. 

\appendix

\section{Lessons from local extensions of $R^2$ inflation}
\label{App:LR2in}
Higher curvature modification of $R^2$ inflation is motivated both from quantum gravity and phenomenological point of view \cite{Ferrara:2013rsa,Asaka:2015vza}.
Higher curvature extensions of $R^2$ model with $R^n$ terms such as 
\begin{equation}
	S_{R^3}= \frac{1}{2}\int d^4x\sqrt{-g}\LT M_p^2R +f_0  \left(R^2 + \frac{f_1}{6 M^2} R^3 \right) \RT 
	\label{r3r4ac}
\end{equation}
\begin{equation}
	S_{R^4}= \frac{1}{2}\int d^4x\sqrt{-g}\LT M_p^2R +f_0  \left(R^2  +\frac{f_2}{4M^4} R^4\right) \RT
	\label{r3r4ac1}
\end{equation}
were extensively studied in recent years  \cite{Huang:2013hsb,Ivanov:2021chn,Motohashi:2014tra,Bamba:2015uma,Asaka:2015vza} and  it was found that the dimensionless coefficients of $R^3$ and $R^4$ terms must be  $f_1\lesssim \Oc\LF 10^{-4} \RF$ and $f_2\lesssim \Oc\LF 10^{-7} \RF$ (reading from \cite{Ivanov:2021chn}) to be compatible with the present constraints on $\LF n_s,\,r\RF$. This conveys that any higher curvature correction beyond $R^2$ term should not modify significantly the background evolution during $N\sim (50-60)$ e-folds from the end of inflation, otherwise we severely spoil the approximate scale invariance of the observed power spectrum of scalar perturbations and especially the prediction $n_s = 1-\frac{2}{N}$, so that the model will no longer be compatible with the CMB data. Furthermore, the effects of these higher curvature corrections are expected to be even more negligible after the end of inflation, i.e., at the stage of reheating when the $R$ and $R^2$ terms dominate. Similar conclusions can be drawn with other higher curvature modifications such as $R^{3/2}$~\cite{Martins:2020oxv} and those inspired from the asymptotic safety (AS) approach. In the latter case, suitable renormalization group (RG) based re-summation of quantum corrections to $R^2$ term gives~\cite{Demmel:2015oqa}
\begin{equation}
	S_{\rm AS}= \frac{1}{2}\int d^4x\sqrt{-g}\Bigg[ M_p^2R+ \frac{aR^2}{1+b\ln\LF \frac{R}{\mu}\RF}  \Bigg]\,.
\end{equation}
From the conclusions of \cite{Liu:2018hno} where inflationary observables for this action were calculated, we learn that if $b\ll 10^{-3}$ the inflationary predictions are very close to those of $R^2$ model, whereas for $b\geq 10^{-3}$ predictions slightly deviate and the tensor-to-scalar ratio can be as large as $r\sim 10^{-2}$ which can be verified in the future detection of B-modes \cite{CMB-S4:2020lpa}. Apart from $\LF n_s,\,r  \RF$, the running of spectral index is also a crucial parameter that gets affected by deviations from $R^2$ model (see \cite{Liu:2018hno} for details). Currently the running of spectral index is severely constrained: 
\begin{equation}
	\frac{dn_s}{d\ln k}\Bigg\vert_{k=k_\ast} = -0.0045\pm 0.0067\,\,at\,\,68\%\,{\rm CL}\,. 
\end{equation}
So, future CMB probes and even more precise measurement of $n_s$ will tell us more about the effects of higher curvature terms. 

Instead of higher curvature extensions, higher derivative (6th order) corrections to the $R^2$-inflation studied in \cite{Castellanos:2018dub} with the following action 
\begin{equation}
	S^6_{R^2} = \frac{M_p^2}{2}\int \Bigg[ R+\alpha R^2+\gamma R\square R \Bigg]
	\label{6thorg}
\end{equation}
where the $6^{\rm th}$ order term $R\square R$ is considered as a small perturbation with $\beta=\frac{\gamma}{6\alpha^2}\ll 1 $. The inflationary predictions $\LF n_s,\,r \RF$ \eqref{6thorg} were estimated within the background solution $\bar{\square} \bar{R}\approx M^2\square{R}$ as
\begin{equation}
	n_s = 1-\frac{2}{N}\pm\frac{\beta}{N},\quad r= \frac{12}{N^2}\pm\frac{\beta}{N^2}\,,
\end{equation}
where $\vert\beta\vert\ll 1$ and its maximum value is $
\vert\beta\vert_{\rm max}=0.3$ inferred from the Planck 2018 data \cite{Castellanos:2018dub}. Moreover, studies of reconstructing f(R) based on the Planck data of $\LF n_s,\,r \RF$ \cite{Sebastiani:2015kfa} also revealed that $R^2$ inflation is the best fit paradigm once we assume the number of e-foldings to be $N=55-60$.  
Therefore, from all this knowledge of geometrical modifications of $R^2$ theory we can fairly assume that  observationally compatible FLRW background must be very close to the solution of the eigenvalue equation \eqref{mR2in}. 

\section{Equations of motion of \eqref{NAID}}
\label{App:EoM}

In this appendix we work out the equations of motion (EOMs) of our model \eqref{NAID}.
First, we reiterate the action in a bit more general way as follows. EOMs are obtained by computing
\begin{equation}
	\delta S_H^{\rm Non-local}=\int d^4x\frac{\delta(\sqrt{-g}L)}{\delta g_{\mu\nu}}h_{\mu\nu}=\int d^4x\frac{E^{\mu\nu}}2h_{\mu\nu}
	\label{EOMdef}
\end{equation}
We start by deriving the part of the EOMs coming from terms quadratic in Weyl keeping only the contributions where one of the Weyl tensors is varied. These terms are relevant for the quadratic variation and derivation of the propagator.
Using the explicit expression for the Weyl tensor in 4 dimensions
\begin{equation*}
	W^{\mu\alpha}_{\phantom{\alpha}\nu\beta}= R^{\mu\alpha}_{\phantom{\alpha}\nu\beta} -\frac 12(\delta^\mu_\nu R^\alpha_{\beta}-\delta^{\mu}_\beta R^{\alpha}_{\nu}+R^\mu_\nu{\delta}^{\alpha}_{\beta}-R^\mu_\beta{\delta}^\alpha_\nu
	)+\frac R{6} (\delta^\mu_\nu {\delta}^{\alpha}_{\beta}-\delta^{\mu}_\beta {\delta}^{\alpha}_{\nu})
\end{equation*}
which uses the position of indexes such that explicit metrics are avoided, one can compute
\begin{equation*}
	\begin{split}
		&\delta (\Oc_1R\Oc_2W^{\mu\alpha}_{\phantom{\alpha}\nu\beta}\Oc_3W_{\mu\alpha}^{\phantom{\mu\alpha}\nu\beta})\\
		=&\Oc_1R\left(\Oc_2\delta R^{\mu\alpha}_{\phantom{\alpha}\nu\beta}\Oc_3W_{\mu\alpha}^{\phantom{\mu\alpha}\nu\beta}+\Oc_2W^{\mu\alpha}_{\phantom{\alpha}\nu\beta}\Oc_3\delta R_{\mu\alpha}^{\phantom{\mu\alpha}\nu\beta}\right)+O(W^2)\\
		=&h^\mu_\beta(R_{\alpha\nu}+2\nabla_\alpha\nabla_\nu)\left[\Oc_2(\Oc_1R\Oc_3W_\mu^{\phantom{\mu}\alpha\nu\beta})+\Oc_3(\Oc_1R\Oc_2W_\mu^{\phantom{\mu}\alpha\nu\beta})\right]+O(W^2)
	\end{split}
\end{equation*}
From the above derivation we can deduce 
\begin{equation*}
	\delta(W_{\mu\nu\rho\sigma}\Oc\LF \square_s,\, R_s \RF W^{\mu\nu\rho\sigma})=h^\mu_\beta2(R_{\alpha\nu}+2\nabla_\alpha\nabla_\nu) \Oc\LF \square_s,\, R_s \RF W_\mu^{\phantom{\mu}\alpha\nu\beta}+O(W^2)
\end{equation*}
The part of EOMs linear in Weyl tensor thus reads (we use twice variation of the action)
\begin{equation}
	{E^\beta_\mu}_W=(R_{\alpha\nu}+2\nabla_\alpha\nabla_\nu)\left\{2\Fc_W W_\mu^{\phantom{\mu}\alpha\nu\beta}+2f_0\frac{\lambda_R}{\Mc_s^2}\left[\Oc_2(\Oc_1R\Oc_3W_\mu^{\phantom{\mu}\alpha\nu\beta})+\Oc_3(\Oc_1R\Oc_2W_\mu^{\phantom{\mu}\alpha\nu\beta})\right]\right\}
	\label{EOMW}
\end{equation}
We notice that terms linear in Weyl in EOMs would vanish in the trace of Einstein equations due to the total tracelessness of the Weyl tensor.

Turning to the part containing only $R$, we start by noting that
\begin{equation*}
	\int d^4x\sqrt{-g}(\delta(A\Oc(\Box_s)B))=\int d^4x\sqrt{-g}\left(\delta A\Oc\LF \square_s \RF B+A\Oc\LF \square_s \RF \delta B+\sum_{n\geq1}f_n\sum_{l=0}^{n-1}A^{(l)}\delta\Box_sB^{(n-1-l)}\right)
\end{equation*}
as long as $\Fc=\sum\limits_{n\geq0}f_n\Box_s^n$ and total derivatives are omitted. Also $A^{(n)}\equiv \Box_s^n A$.
d'Alembertian acting on scalars is simply given $g^{\mu\nu}\nabla_\mu\pd_\nu$ and its variation reads
\begin{equation*}
	\delta(\Box_s)B=\frac1{\Mc_s^2}(-h^{\mu\nu}\nabla_\mu\pd_\nu-g^{\mu\nu}\gamma^\rho_{\mu\nu}\pd_\rho)B
\end{equation*}
The double summation term upon moving derivatives transforms into
\begin{equation*}
	\int d^4x\sqrt{-g}\left(\frac{h^\mu_\beta}{\Mc_s^2}\sum_{n\geq1}f_n\sum_{l=0}^{n-1}\left[\pd^\beta A^{(l)}\pd_\mu B^{(n-l-1)}-\frac12\delta^\beta_\mu\pd^\nu A^{(l)}\pd_\nu B^{(n-l-1)}-\frac12\delta^\beta_\mu A^{(l)}\Box B^{(n-l-1)}\right]\right)
\end{equation*}
When $A=B=R$ and $\Oc=\Fc_R\LF \square_s \RF$, we have
\begin{equation}
	{E^\beta_\mu}_{R^2}=2(-R^\beta_\mu+\D^\beta\D_\mu-\delta^\beta_\mu\Box)\Fc_R R+\Kc^\beta_\mu-\frac12\delta^\beta_\mu\Kc^\nu_\nu-\frac12\delta^\beta_\mu\tilde{\Kc}
	\label{EOMR2}
\end{equation}
where we have defined
\begin{equation*}
	\Kc_{\mu}^{\beta} =  \frac{1}{\Mc_{s}^{2}}\sum_{n=1}^{\infty}f_{Rn}\sum_{l=0}^{n-1}\partial^{\beta}R^{(l)}\partial_{\mu}R^{(n-l-1)}\,,\quad
	\tilde{\Kc}=   \sum_{n=1}^{\infty}f_{Rn}\sum_{l=0}^{n-1}R^{(l)}R^{(n-l)}
\end{equation*}
Dealing with the cubic in curvature term when $A=\Lc_1\LF \square_s \RF R\Lc_2\LF \square_s \RF R$, $B=R$ and $\Oc=\Lc_3\LF \square_s \RF$, we have
\begin{equation}
	\delta(\Lc_1R\Lc_2R)\Lc_3R+(-R^\beta_\mu+\D^\beta\D_\mu-\delta^\beta_\mu\Box)\Lc_3(\Lc_1R\Lc_2R)+{\Kc_3}_\mu^\beta
	-\frac12\delta^\beta_\mu{\Kc_3}_\nu^\nu
	-\frac12\delta^\beta_\mu{\tilde{\Kc}_3\mu}^\beta 
	\label{R33}
\end{equation}
where we have defined
\begin{equation*}
	{\Kc_3}_{\mu}^{\beta} =  \frac{1}{\Mc_{s}^{2}}\sum_{n=1}^{\infty}l_{3n}\sum_{l=0}^{n-1}\partial^{\beta}(\Lc_1R\Lc_2R)^{(l)}\partial_{\mu}R^{(n-l-1)}\,,\quad
	\tilde{\Kc}_3=   \sum_{n=1}^{\infty}l_{3n}\sum_{l=0}^{n-1}(\Lc_1R\Lc_2R)^{(l)}R^{(n-l)}
\end{equation*}
and the first term in (\ref{R33}) results in a cycling rotation of indices. The result with all normalizations and the factor $2$ reads
\begin{equation}
	{E^\beta_\mu}_{R^3}=2 \frac{\lambda_3}{\Mc_s^2}\left[(-R^\beta_\mu+\nabla^\beta\nabla_\mu-\delta^\beta_\mu\Box)\Rc_3+\sum_{i=1,2,3}\left({\Kc_i}_\mu^\beta-\frac12\delta^\beta_\mu{\Kc_i}_\nu^\nu-\frac12\delta^\beta_\mu{\tilde\Kc_i}\right)\right]
	\label{EOMR3}
\end{equation}
where $$\Rc_3=\Lc_1(\Lc_2R\Lc_3R)+\Lc_2(\Lc_1R\Lc_3R)+\Lc_3(\Lc_1R\Lc_2R)$$

The remaining is the Einstein term and the square root of $-g$ variation canonically leading to
\begin{equation}
	{E_\mu^\beta}_R=\delta^\beta_\mu L_H^{\rm Non-local}-M_P^2R^\beta_\mu
	\label{EOMR}
\end{equation}

Performing the routine one has
\begin{equation}
	\begin{aligned}
		E^\beta_\mu=&\delta^\beta_\mu L_H^{\rm Non-local}-M_P^2R^\beta_\mu
		+2(-R^\beta_\mu+\D^\beta\D_\mu-\delta^\beta_\mu\Box)\Fc_R R+\Kc^\beta_\mu-\frac12\delta^\beta_\mu\Kc^\nu_\nu-\frac12\delta^\beta_\mu\tilde{\Kc}\\
		+&\frac23\frac{\lambda_c}{\Mc_s^2}\left[(-R^\beta_\mu+\D^\beta\D_\mu-\delta^\beta_\mu\Box)\Rc_3+\sum_{i=1,2,3}\left({\Kc_i}_\mu^\beta-\frac12\delta^\beta_\mu{\Kc_i}_\nu^\nu-\frac12\delta^\beta_\mu{\tilde\Kc_i}\right)\right]\\
		+&(R_{\alpha\nu}+2\nabla_\alpha\nabla_\nu)\left\{2\Fc_W W_\mu^{\phantom{\mu}\alpha\nu\beta}+\frac23\frac{\lambda_R}{\Mc_s^2}\left[\Dc_2(\Dc_1R\Dc_3W_\mu^{\phantom{\mu}\alpha\nu\beta})+\Dc_3(\Dc_1R\Dc_2W_\mu^{\phantom{\mu}\alpha\nu\beta})\right]\right\}+\cdots
	\end{aligned}
	\label{NAIDeq}
\end{equation}
where auxiliary definitions from the above are used and $\cdots$ contain terms irrelevant for both the FLRW background and linear perturbations because all these additional terms $\cdots$ are quadratic in Weyl tensor. 
Computing the trace, one gets
\begin{equation}
	\begin{split}
		E=&{M_{p}^{2}}R +
		\frac43\frac{\lambda_3}{\Mc_s^2}\Lc_1\LF \square_{s} \RF R\, \Lc_2\LF \square_{s} \RF R\, \Lc_3\LF  \square_{s}\RF R
		-6\Box\Fc_RR-(\Kc_\nu^\nu+2\tilde\Kc)\\
		-&2\frac{\lambda_c}{\Mc_s^2}\left[(R+3\Box)\Rc_3+\sum_{i=1,2,3}\left({\Kc_i}_\nu^\nu+2{\tilde\Kc_i}\right)\right]+\dots
	\end{split}
	\label{NAIDappEOMtrace}
\end{equation}
where we see that linear in Weyl tensor terms went away identically due to the tracelessness of the Weyl tensor and $\cdots$ are the terms irrelevant even for linear variations. We remind that in the absence of matter, solving the trace equation on a spatially-flat FLRW background is enough to solve all the system modulo perhaps some radiation source. As such we start solving the EOMs with the latter equation as it seems to be the simplest one, even though highly non-trivial. Now we try to see whether Starobinsky solution \eqref{mR2in} can be a solution here. This would preserve a lot of the famous local $R^2$ inflation model. Substituting 
this into the EOMs simplifies them considerably. First we compute
\begin{equation}
	\Kc_{\mu}^{\beta} =  \frac{1}{\Mc_{s}^{2}}\Fc_R^\dagger\left(r_s\right)\partial^{\beta}R\partial_{\mu}R\,,\quad
	\tilde{\Kc}=  r_s\Fc_R^\dagger\left(r_s\right)R^2
\end{equation}
\begin{equation*}
	\begin{aligned}
	{\Kc_3}_{\mu}^{\beta} & =  \frac{1}{\Mc_{s}^{2}}\Lc_1(r_s)\Lc_2(r_s)\pd^\beta\left(\frac{\Lc_3(\Box_s)-\Lc_3(r_s)}{\Box_s-r_s}R^2\right)\partial_{\mu}R \\ 
	\tilde{\Kc}_3 & =   r_s\Lc_1(r_s)\Lc_2(r_s)\left(\frac{\Lc_3(\Box_s)-\Lc_3(r_s)}{\Box_s-r_s}R^2\right)R
	\end{aligned}
\end{equation*}
where $r_s= M^2/\Mc_s^2$. Now we, can notice that (\ref{mR2in}) passes through all equations proving it is a solution as long as FLRW backgrounds are considered (meaning that Weyl tensor vanishes) and conditions (\ref{condiq}) are met.

\section{Degrees of freedom and two point correlations of \eqref{NAID} in quasi-dS limit}
\label{App:pEoM}

In this section we compute the inflationary observables related to two point correlations of fluctuations. To study quantum fluctuations generated during inflation, we first define metric perturbations by the following line element  in terms of the gauge invariant Bardeen potentials $\left(\Phi,\,\Psi\right)$ and the transverse and traceless tensor fluctuations $h_{ij}$ 
\begin{equation}
	\begin{aligned}
		ds^{2}=a^{2}\left(\tau\right)\Bigg[-&\left(1+2\Phi\right)d\tau^{2} +\left(\left(1-2\Psi\right)\delta_{ij} +2h_{ij}\right)dx^{i}dx^{j}\Bigg]\,.\label{line-element}
	\end{aligned}
\end{equation}
The above perturbed metric remains the same when we choose the Newtonian gauge. 

By studying the linearized perturbed equations \eqref{NAIDeq} in the dS approximation, i.e. assuming $\bar{ R}\approx {\rm constant}$, we obtain
\begin{equation}
	\Bigg[ f_0\bar{ R}_{\rm dS}+ f_0\bar{ R}_{\rm dS}\LF \bar{\square}_{\rm dS}-\frac{\bar{ R}_{\rm dS}}{6} \RF \Fc_{R}\LF \frac{\bar{\square}_{\rm dS}}{\Mc_s^2}+\frac{\bar{ R}_{\rm dS}}{2\Mc_{s}^2} \RF\Bigg]\frac{\Phi+\Psi}{a^2} = 0. 
	\label{phisi}
\end{equation}
Substituting the form-factor \eqref{formfinal} in \eqref{phisi}, 
we obtain
\begin{equation}
	e^{\gamma_T\LF \frac{\bar{\square}_{\rm dS}}{\Mc_s^2}-\frac{\bar{R}_{\rm dS}}{6\Mc_s^2} 
		\RF}\LF \Phi+\Psi  \RF =0\,,
	\label{phisi0}
\end{equation}
whose general solution is $\Phi+\Psi=0$ since the exponent of an entire function of d'Alembertian operator do not generate any new solutions that is extensively explored in \cite{Koshelev:2017tvv}. At the second order level around dS, the part of the action \eqref{NAID} containing Weyl tensor terms becomes equivalent to the one studied in \cite{Koshelev:2016xqb,Koshelev:2017tvv,Koshelev:2020foq}. Therefore, our result \eqref{phisi0} matches with Eq. 5.33 of \cite{Koshelev:2016xqb}. We, however, show this explicitly in Appendix~\ref{App:pEoM}. 

Imposing \eqref{phisi} it is straight forward to see from \eqref{scalwper} that the scalar part of the Weyl tensor perturbation vanishes \cite{Koshelev:2016xqb}
\begin{equation}
	\delta_{(s)} W^{\alpha\beta}_{\quad \mu\nu} = \frac{\Phi+\Psi}{a^2} K^{\alpha\beta}_{\quad \mu\nu}\,, 
	\label{scalwper}
\end{equation}
where
\begin{equation}
	\begin{aligned}
		K^{0i}_{\quad 0j} & = -\frac{1}{6}\delta^i_jk^2+\frac{1}{2}k^ik_j \\
		K^{il}_{\quad jm} &= \frac{1}{3}k^2\LF \delta^l_m\delta^i_j -\delta^l_j\delta^i_m \RF -\frac{1}{2}\delta^l_m k^ik_j -\frac{1}{2}\delta^i_j k^l k_m + \frac{1}{2} \delta^l_j k^ik_m+\frac{1}{2}\delta^i_m k^lk_j\,.
	\end{aligned}
\end{equation}
Thus, the second order action for the scalar perturbations for \eqref{NAID} become 
\begin{equation}
	\delta_{(s)}^{(2)}S = \frac{1}{2f_0\bar{R}_{\rm dS}} \int d^4x\sqrt{-\bar{g}} \Upsilon\frac{\Wc\LF\bar{\square}_s\RF}{\Fc_R\LF \bar{\square}_s \RF}\LF \bar{\square}_{\rm dS} -M^2 \RF \Upsilon\,. 
	\label{Upsiloneq}
\end{equation}
where $\Upsilon = 2f_0\bar{R}_{\rm dS}\Psi$ is the canonical variable, and it is related to the curvature perturbation as $\Upsilon\approx -2f_0\bar{R}_{\rm dS}\Rc$. The operator $\Wc\LF \square_s \RF$ is related to the form-factor $\Fc_R\LF \square_s \RF$ as  
\begin{equation}
	\Wc\LF \square_s \RF = 3\Fc_R\LF \square_s \RF + \LF \bar{R}_{\rm dS}+3M^2 \RF \frac{\Fc_R\LF \square_s \RF-f_0}{\square-M^2}\,,
	\label{opW}
\end{equation}
From \eqref{Upsiloneq} we can first deduce that the kinetic term of $\Upsilon$ has one real zero corresponding to  $\bar{\square}_{\rm dS} = M^2$ which indicates there is one propagating degree of freedom which we call `scalaron'. 
If there are any other degrees of freedom, they must arise from zeros of the operator $\Wc\LF \square_s \RF$ \eqref{opW}.\footnote{If $\Wc\LF \square_s \RF =3f_0 e^{\gamma_0\LF \square_s+\frac{\bar{R}_{\rm dS}}{3\Mc_s^2} \RF} $, where $\gamma_0$ is an entire function of $\square_s+\frac{\bar{R}_{\rm dS}}{3\Mc_s^2}$ operator, 
	then it will have no zeros in the entire complex plane. But as a consequence the form-factor $\Fc_R\LF \square_s \RF$ must depend on the background $\bar{R}_{\rm dS}$ as \cite{Craps:2014wga,Koshelev:2020foq}
	\begin{equation}
		\Fc_R\LF \square_s\RF \equiv  \Fc_R\LF\square_s,\,\bar{R}_{\rm dS}\RF =  f_0\frac{e^{\gamma_0\LF \square_s+\frac{\bar{R}_{\rm dS}}{3\Mc_s^2}\RF} \LF \square_s-\frac{M^2}{\Mc_s^2} \RF+\LF \bar{R}_{\rm dS}+3M^2 \RF }{3\square+\bar{R}_{\rm dS}}
		\label{formds}
	\end{equation}
	This choice of form-factor was considered in \cite{Craps:2014wga,Koshelev:2020foq}, but it leads to background dependence in the action \eqref{NAID} through the factor $\bar{R}_{\rm dS}$ in \eqref{formds}. In this paper we rather consider the choice of form-factor $\Fc_R\LF \square_s \RF$ corresponding to \eqref{formMin}, so that we can take a smooth low-energy limit and avoid background dependence. }  Computing it for the form-factor \eqref{formMin}, we obtain 
\begin{equation}
	\Wc\LF \square_s \RF = -f_0\bar{R}_{\rm dS} \LF \frac{1-e^{\gamma_S\LF \square_s \RF}}{\square} \RF + 3f_0 e^{\gamma_S\LF \square_s \RF} 
	\label{WFMin}
\end{equation}

To obtain zeros of $\Wc\LF \square_s \RF$, we need to solve the following characteristic equation 
\begin{equation}
	\Wc\LF \frac{Z}{\Mc_s^2} \RF = 0 \implies \bar{R}_{\rm dS}\LF \frac{1-e^{\gamma_S\LF  \frac{Z}{\Mc_s^2} \RF}}{Z} \RF = 3e^{\gamma_S\LF  \frac{Z}{\Mc_s^2} \RF} \,.
	\label{characeq}
\end{equation}
To solve \eqref{characeq}, first we must specify the choice of entire function $\gamma_S\LF \square_s \RF$. For simplicity, let us take  a compatible choice of entire function \eqref{chformFR} corresponding to $p_i\LF \square_s \RF = \alpha_1\square_s$ that gives
\begin{equation}
	\gamma_S\LF \square_s \RF = \alpha_1 \square_s\LF \square_s-\frac{M^2}{\Mc_s^2} \RF\,. \label{chformFR}
\end{equation}
Substituting \eqref{chformFR} into the characteristic equation \eqref{characeq}, we obtain no real solutions but instead we get infinite tower of complex conjugate solutions given by
\begin{equation}
	\begin{aligned}
		Z = &\, \frac{M^2}{2}+\frac{\Mc_s^2}{2}\LT \LF 2\pi+4q\pi\RF^2+\LF\frac{M}{\Mc_s}\RF^8 \RT^{1/4}\Bigg\{\cos\LT \frac{1}{2}{\rm Arg}\LT 4\pi i\LF q+\frac{1}{2}\RF+\frac{M^4}{\Mc_s^4} \RT \RT \\& + i \sin\LT \frac{1}{2}{\rm Arg}\LT 4\pi i\LF q+\frac{1}{2}\RF+\frac{M^4}{\Mc_s^4} \RT\RT\Bigg\}\\ 
		\approx &\,\Bigg\vert_{M^2\ll \Mc_s^2}  \pm \Mc_s^2 \sqrt{q+\frac{1}{2}}\LF 1 \pm i\RF
	\end{aligned}
	\label{ccstates}
\end{equation}
where $q\geq 1$ is a positive integer. In the context of string field theory, it was known that complex conjugate poles give (classical) degrees of freedom for a coupled system of scalar fields (both with positive and negative kinetic terms in equal numbers) \cite{Koshelev:2007fi,Koshelev:2009ty,Koshelev:2010bf,Arefeva:2008zru}. The primary question here is whether degrees of freedom corresponding to the infinite tower of complex conjugate poles are physical and contribute to the inflationary correlations. 
In this context, we can take inspiration from the studies in different context \cite{Buoninfante:2018lnh,Buoninfante:2020ctr} where non-local scalar field theories with infinitely many complex conjugate poles are considered with the following type of Lagrangians 
\begin{equation}
	\Lc_{\phi} = \frac{1}{2}\phi \LF e^{\gamma_\phi\LF \square_s \RF}-1 \RF \phi -V(\phi)\,, 
	\label{scLag}
\end{equation}
where $\gamma_\phi$ being an arbitrary entire function. It was shown that the Lagrangians of this type satisfy the  optical theorem both at the tree-level \cite{Buoninfante:2018lnh,Buoninfante:2020ctr} and one-loop level \cite{Luca} due to the exact cancellation of contributions from complex conjugate poles. 
Unitarity due to the presence of complex conjugate poles has been studied in the context of Lee-Wick theories, and it was proposed that one can project away the states corresponding to complex conjugate poles using new quantum field theory prescriptions (such as fakeon). Thus, one can disregard these states as unphysical\footnote{One can alternatively chose a suitable entire function such that the imaginary part of the pole is small enough to avoid any classical instabilities \cite{Koshelev:2021orf}. But however, such a choice of entire function necessarily should depend on the background value of $\bar{R}_{\rm dS}$. In the present construction we aim to avoid any background dependence of form-factors.} \cite{Modesto:2015ozb,Modesto:2016ofr,Anselmi:2017lia,Anselmi:2017ygm,Anselmi:2018kgz,Anselmi:2020lpp,Anselmi:2021hab,Anselmi:2022toe,Liu:2022gun,Frasca:2022gdz}. 

All the above mentioned studies are about the non-local scalar field theories in Minkowski space-time. Our case is in the context of gravity and our gravitational action \eqref{NAID} has no degrees of freedom (classically) with complex conjugate poles around Minkowski space-time. Indeed, we can observe that in the Minkowski limit we have $\Wc\LF \square_s \RF\to 3f_0 e^{\gamma_S\LF  \square_s\RF}$ (see \eqref{WFMin}), and therefore we have only one real pole for the kinetic operator of $\Upsilon$ at $\bar{\square} = M^2$. This is expected as our form-factor \eqref{formMin} is fixed for the ghost-free condition around Minkowski background \cite{Biswas:2016egy}. It is well known that $R^2$ inflation is an intermediate phase of evolution which is unstable to the past as well as to the future \cite{Starobinsky:1980te,Muller:1987hp,Muller:2017nxg}. Moreover, when we quantize the inflationary modes, we take them  to be deep inside the Hubble radius $k\gg aH$ and impose adiabatic vacuum initial conditions \cite{Mukhanov:1990me,Koshelev:2020foq}. The Fourier modes deep inside the Hubble radius $k\gg aH$ are the ones close to the locally Minkowski space-time since the effects of dS curvature can be neglected \cite{Mukhanov:1990me}. Therefore, physically for the purposes of inflationary quantum fluctuations, we ignore all the complex conjugate modes \eqref{ccstates} by setting initial conditions of them to be zero deferring the detailed study of non-local (quantum) theories in curved space-time for future investigations. 

Thus, in addition to massless graviton, we have only one propagating scalar (scalaron) with mass $M^2$ in Minkowski space-time. Thus, we can deduce from \eqref{Upsiloneq} that solutions of $\Upsilon$ are governed by
\begin{equation}
	\LF \bar{\square}_{\rm dS}-M^2 \RF \Upsilon = 0\,,
	\label{Upsilonsol}
\end{equation}
Following the previous studies \cite{Craps:2014wga,Koshelev:2017tvv}, we redefine the field $\Upsilon$ after imposing the on-shell condition \eqref{Upsilonsol} as $\Upsilon\to \sqrt{\frac{\Wc\LF \frac{M^2}{\Mc_s^@} \RF}{f_0\bar{R}_{\rm dS}\Fc_R\LF \frac{M^2}{\Mc_s^2} \RF}}\Upsilon$. As a result, we obtain the second order action for the curvature perturbation $	\Rc = \Psi +H\frac{\delta R}{\dot{\bar{	R}}} \approx -\frac{\Psi}{\epsilon}$ identical to that in the local $R^2$-theory:
\begin{equation}
	\delta^{(2)}_{(s)} S_{H}^{Non-local} = \frac{M_p^2}{2}\epsilon\int d^4x\sqrt{-\bar{	g}}\, \Rc\LF \bar{	\square}_{\rm dS}-M^2 \RF\Rc\, ,
	\label{s2s}
\end{equation}
where $\bar{\square}_{\rm dS}$ is the d'Alembertian operator in the dS approximation. 

\subsection{Tensor fluctuations around $\bar{\square}\bar{R}=M^2\bar{R}$ and dS approximation}

\label{App:tensor-pert}

Computing the second order action of \eqref{NAID} for transverse and traceless tensor perturbations $h_{ij}$ \eqref{line-element} in the dS approximation, i.e., $\bar{R}_{\rm dS}\approx \const$, we get
\begin{equation}
	\begin{aligned}
		\delta^{(2)}_{(h)}S_H^{\rm Non-local} = \,& \int d^4x\sqrt{-\bar{	g}}\Bigg[\LF\frac{M_p^2}{2}+f_0\bar{	R}_{\rm dS}\RF \LF \delta^{(2)}R +\delta^{(2)}_g\RF -\frac{f_0}{2}\delta^{(2)}_g f_0\bar{	R}_{\rm dS} \\ & + f_0\bar{	R}_{\rm dS}\, \delta W_{\mu\nu\rho\sigma}\Fc_W\LF \frac{\bar{ \square}_{\rm dS}}{\Mc_s^2},\, \frac{\bar{	R}_{\rm dS}}{\Mc_s^2} \RF \delta W^{\mu\nu\rho\sigma} \\ & + \frac{f_0\lambda_R}{\Mc_s^2} \Dc_1\LF \frac{M^2}{\Mc_s^2} \RF \Dc_2\LF \frac{\bar{	\square}_{\rm dS}}{\Mc_s^2} \RF  \delta W_{\mu\nu\rho\sigma}  \Dc_3\LF \frac{\bar{	\square}_{\rm dS}}{\Mc_s^2} \RF\delta W^{\mu\nu\rho\sigma}  \Bigg]\,.
	\end{aligned}
	\label{haction}
\end{equation}
Imposing $\Dc_1\LF \frac{M^2}{\Mc_s^2} \RF =0$ and using $\Fc_W\LF \square_s,\,R_s \RF$ 
from \eqref{formfinal} with the approximation $M_p^2 \ll f_0\bar{	R}_{\rm dS}$,  
the second order action for tensor fluctuations becomes
\begin{equation}
	\delta_{(t)}^{(2)} S_{H}^{\rm Non-local} = \int d^4x \sqrt{-g} \Bigg[h_{ij} \, e^{\gamma_T\LF \frac{\bar{	\square}_{\rm dS}}{\Mc_s^2} -\frac{\bar{ R}_{\rm dS}}{3\Mc_s^2} \RF}\LF\bar{	\square}_{\rm dS} -\frac{\bar{R}_{\rm dS}}{6} \RF h^{ij}\Bigg]
	\label{s2h}
\end{equation}
that coincides with the results of \cite{Koshelev:2016xqb,Koshelev:2020foq}. 

\bibliographystyle{utphys}
\bibliography{ssa.bib}
\end{document}